\documentclass[english,aps,pre,preprint,superscriptaddress,nofootinbib]{revtex4}
\usepackage{amsmath}
\usepackage{babel}
\usepackage{graphics}
\usepackage{amssymb}

\newcommand{\noun}[1]{{\tt#1}}

\begin{document}


\title{A Linear Programming Algorithm to Test for Jamming in Hard-Sphere
Packings}

\author{Aleksandar Donev}

\affiliation{\emph{Program in Applied and Computational Mathematics}, \emph{Princeton
University}, Princeton NJ 08540}

\affiliation{\emph{Princeton Materials Institute}, \emph{Princeton University},
Princeton NJ 08540}

\author{Salvatore Torquato}

\email{torquato@electron.princeton.edu}

\affiliation{\emph{Princeton Materials Institute}, \emph{Princeton University},
Princeton NJ 08540}

\affiliation{\emph{Department of Chemistry}, \emph{Princeton University}, Princeton
NJ 08540}

\author{Frank H. Stillinger}

\affiliation{\emph{Department of Chemistry}, \emph{Princeton University}, Princeton
NJ 08540}

\author{Robert Connelly}

\affiliation{\emph{Department of Mathematics}, \emph{Cornell University}, Ithaca
NY 14853}

\date{\today}

\begin{abstract}
Jamming in hard-particle packings has been the subject of considerable
interest in recent years. In a paper by Torquato and Stillinger {[}J. Phys. Chem.
B, 105 (2001){]}, a classification scheme of jammed packings into
hierarchical categories of \emph{locally}, \emph{collectively} and
\emph{strictly jammed} configurations has been proposed. They suggest
that these jamming categories can be tested using numerical algorithms
that analyze an equivalent contact network of the packing under applied
displacements, but leave the design of such algorithms as a future
task. In this work we present a \emph{rigorous} and \emph{efficient}
algorithm to assess whether a hard-sphere packing is jammed according
to the afformentioned categories. The algorithm is based on linear
programming and is applicable to regular as well as random packings
of finite size with hard-wall and periodic boundary conditions. If
the packing is not jammed, the algorithm yields representative multi-particle
unjamming motions. We have implemented this algorithm and applied
it to ordered lattices as well as random packings of disks and spheres
under periodic boundary conditions. Some representative results for
ordered and disordered packings are given, but more applications are
anticipated for the future. One important and interesting result is
that the random packings that we tested were strictly jammed in three
dimensions, but \emph{not} in two dimensions. Numerous interactive
visualization models are provided on the authors' webpage. 
\end{abstract}
\maketitle

\section{introduction}

\newcommand{\Vr}{\mathbf{r}}

\newcommand{\Vdr}{\Delta \mathbf{r}}

\newcommand{\VR}{\mathbf{R}}

\newcommand{\VdR}{\Delta \mathbf{R}}

\newcommand{\Vv}{\mathbf{v}}

\newcommand{\VV}{\mathbf{V}}

\newcommand{\Vu}{\mathbf{u}}

\newcommand{\Vnc}{\mathbf{n}_{c}}

\newcommand{\Vb}{\mathbf{b}}

\newcommand{\Vf}{\mathbf{f}}

\newcommand{\Vl}{\mathbf{l}}

\newcommand{\Vdl}{\Delta \mathbf{l}}

\newcommand{\VNc}{\mathbf{N}_{c}}

\newcommand{\MA}{\mathbf{A}}

\newcommand{\Mepsilon}{\boldsymbol {\varepsilon }}

\newcommand{\Vlambda}{\boldsymbol {\lambda }}

\newcommand{\MLambda}{\boldsymbol {\Lambda }}

\newcommand{\MdLambda}{\Delta \boldsymbol {\Lambda }}

Packings of hard particles interacting only with infinite repulsive
pairwise forces on contact are applicable as models of complex manybody
systems because repulsive interactions are the primary factor in determining
their structure. Hard-particle packings are therefore widely used
as simple models for granular media \cite{Granular_Matter,Tensorial_contact},
glasses \cite{Amorphous_solids}, liquids \cite{Simple_Liquids},
and other random media \cite{Random_Materials}, to mention a few
examples. Furthermore, hard-particle packings, and especially hard-sphere
packings, have inspired mathematicians and been the source of numerous
challenging (many still open) theoretical problems \cite{Perfect_packing}.
Of particular theoretical and practical interest are \emph{jammed}
configurations of hard particles. The statistical physics of large
jammed systems of hard particles is a very active field of theoretical
research. For packings of smooth hard spheres a rigorous mathematical
foundation can be given to the concept of jamming, though such rigor
is often lacking in the current physical literature.

There are still many important and challenging questions open even
for the simplest type of hard-particle packings, i.e., monodisperse
packing of smooth perfectly impenetrable spheres. One category of
open problems pertains to the enumeration and classification of disordered
disk and sphere packings, such as the precise identification and quantitative
description of the maximally random jammed (MRJ) state \cite{Torquato_MRJ},
which has supplanted the ill-defined ``random close packed'' state.
Others pertain to the study of ordered systems and finding packing
structures with extremal properties, such as the lowest or highest
(for polydisperse packings) density jammed disk or sphere packings,
for the various jamming categories described below \cite{Stillinger_lattices,Regular_figures}.
Numerical algorithms have long been the primary tool for studying
random packings quantitatively. In this work we take an important
step toward future studies aimed at answering the challenging questions
posed above by designing tools for algorithmic assesment of the jamming
category of finite packings.

In the first part of this paper, we present the conceptual theoretical
framework underlying this work. Specifically, we review and expand
on the hierarchical classification scheme for jammed packings into
locally, collectively and strictly jammed packings proposed in Ref.
\cite{Torquato_jammed}. In the second part, we present a randomized
linear programming algorithm for finding unjamming motions within
the approximation of small displacements, focusing on periodic boundary
conditions. Finally, in the third part we apply the algorithm to monodisperse
packings under periodic boundary conditions, and present some representative
but non-exhaustive results for several periodic ordered lattice packings
as well as random packings obtained via the Lubachevsky-Stillinger
packing algorithm \cite{LS_algorithm}.

Ideas used here are drawn heavily from literature in the mathematical
community (\cite{Connelly_disks,Connelly_packings,Connelly_Tensegrities,Rigidity_theory},
etc.), and these have only recently percolated into the granular materials
community (\cite{Rigidity_theory,Roux,Moukarzel_isostatic}, etc.).
A separate paper \cite{Second_paper} will attempt to give a unified
and more rigorous presentation of the mathematical ideas underlying
the concepts of stability, rigidity and jamming in sphere packings.
Nonetheless, some mathematical preliminaries are given here, a considerable
portion of which is in the form of footnotes.

\subsection{Jamming in Hard-Sphere Packings}

The term jamming has been used often with ambiguity, even though the
physical intuition behind it is strong: It imparts a feeling of being
frozen in a given configuration. Two main approaches can be taken
to defining jamming, \textbf{kinematic} and \textbf{static}. In the
kinematic approach, one considers the motion of particles away from
their current positions, and this approach is for example relevant
to the study of flow in granular media%
\footnote{In particular, the cessation of flow as jamming is approached.
}. The term \emph{jammed} seems most appropriate here. In the static
approach, one considers the mechanical properties of the packing and
its ability to resist external forces%
\footnote{In particular, the infinite elastic moduli near jamming.
}. The term \emph{rigid} is often used among physicists in relation
to such considerations.

However, due to the correspondence between kinematic and static properties,
i.e. strains and stresses, these two different views are largely equivalent.
A more thorough discussion of this duality is delayed to a later work
\cite{Second_paper}, but is touched upon in section \ref{Duality}.
In this paper we largely adopt a kinematic approach, but the reader
should bear in mind the inherent ties to static approaches.

\subsection{Three Jamming Categories}

First we repeat, with slight modifications as in Ref. \cite{Cones_strain},
the definitions of several hierarchical jamming categories as taken
from Ref. \cite{Torquato_jammed}, and then make them mathematically
specific and rigorous for several different types of sphere packings:\\
A finite system of spheres is:

\begin{description}
\item [Locally~jammed]Each particle in the system is locally trapped by
its neighbors, i.e., it cannot be translated while fixing the positions
of all other particles.
\item [Collectively~jammed]Any locally jammed configuration in which no
subset of particles can simultaneously be displaced so that its members
move out of contact with one another and with the remainder set. An
equivalent definition is to ask that all finite subsets of particles
be trapped by their neighbors.
\item [Strictly~jammed]Any collectively jammed configuration that disallows
all globally uniform volume-nonincreasing deformations of the system
boundary. Note the similarity with collective jamming but with the
additional provision of a deforming boundary. This difference and
the physical motivations behind it should become clearer in section
\ref{Strict_Jamming}.
\end{description}
Observe that these are ordered \emph{hierarchically}, with local being
a prerequisite for collective and similarly collective being a prerequisite
for strict jamming.

The precise meaning of some of the concepts used in these definitions,
such as unjamming, boundary and its deformation, depends on the type
of packing one considers and on the particular problem at hand, and
we next make these specializations more rigorous for specific types
of packings of interest. Moreover, we should point out that these
do not exhaust all possibilities and various intricacies can arise,
especially when considering infinite packings, to be discussed further
in Ref. \cite{Second_paper}. It should be mentioned in passing that
jammed random particle packings produced experimentally or in simulations
typically contain a small population of ``rattlers'', i.e., particles
trapped in a cage of jammed neighbours but free to move within the
cage. For present purposes we shall assume that these have been removed
before considering the (possibly) jammed remainder.

\subsection{\label{Unjamming_motions}Unjamming Motions}

Before discussing the algorithms and related issues, it is important
to specify exactly what we mean by unjamming. First, it is helpful
to define some terms. A \textbf{sphere packing} is a collection of
spheres in Euclidian \( d \)-dimensional space \( \Re ^{d} \) such
that the interiors of no two spheres overlap. Here we focus on monodisperse
systems, where all spheres have diameter \( D=2R \), but most of
the results generalize to polydisperse systems as well. A packing
\textbf{\emph{\( P(\VR ) \)}} of \( N \) spheres is characterized
by the arrangement of the sphere centers%
\footnote{Capitalized bold letters will be used to denote \( dN \)-dimensional
vectors which correspond to the \( d \)-dimensional vectors of all
\( N \) of the particles.
} \( \VR =\left( \Vr _{1},\ldots ,\Vr _{N}\right)  \), called \emph{configuration}:
\[
P(\VR )=\left\{ \Vr _{i}\in \Re ^{d},\, \textrm{ }i=1,\ldots ,N\, :\, \left\Vert \Vr _{i}-\Vr _{j}\right\Vert \geq D\, \, \forall j\neq i\right\} \]

The natural physical definition%
\footnote{This is the second definition (definition \emph{b}) in section 2.1
of Ref. \cite{Connelly_Tensegrities}.
} of what it means to unjam a sphere packing is provided by looking
at ways to move the spheres from their current configuration. This
leads us to the definition: An \textbf{unjamming} \textbf{motion}
\( \VdR (t) \), where \( t \) is a time-like parameter, \( t\in \left[ 0,1\right]  \),
is a \emph{continuous} displacement of the spheres from their current
position along the path \( \VR +\VdR (t) \), starting from the current
configuration, \( \VdR (0)=0 \), and ending at the final configuration
\( \VR +\VdR (1) \), while observing all relevant constraints along
the way%
\footnote{This means that impenetrability and any other particular (boundary)
conditions must be observed, i.e. \( P(\VR +\VdR (t)) \) is a \emph{valid}
packing for all \( t\in \left[ 0,1\right]  \).
}, such that some of the contacting spheres lose contact with each
other for \( t>0 \). If such an unjamming motion does not exist,
we say that the packing is \emph{jammed}%
\footnote{Of course by changing the (boundary) constraints we get different
categories of jamming, such as local, collective and strict.
}. It can be shown (see references in Ref. \cite{Connelly_Tensegrities})
that an equivalent definition%
\footnote{This the first definition (definition \emph{a}) in section 2.1 of
Ref. \cite{Connelly_Tensegrities}.
} is to say that a packing is jammed if it is isolated in the allowed
configuration space, i.e., there is no valid packing within some (possibly
small) finite region around \( \VR  \) that are not equivalent%
\footnote{Two packings are equivalent if there is a distance-preserving (rigid-body)
motion (map), such as a uniform translation or a rigid rotation, that
takes one packing to the other.
} to \( P(\VR ) \). We mention this because it is important to understand
that although we use a kinematic description based on motion through
time, which imparts a feeling of dynamics to most physicists, a perfectly
equivalent static view is possible. Therefore, henceforth special
consideration will be given to the final displacement%
\footnote{This will be important when discussing random packings with finite
(but small) interparticle gaps.
} \( \VdR (1) \), so that we will most often just write \( \VdR =\VdR (1) \). 

It is also useful to know that we need to only consider \emph{analytic}
\( \VdR (t) \), which gives us the ability to focus only on derivatives
of \( \VdR (t) \). Furthermore, it is a simple yet fundamental fact%
\footnote{At the second derivative spheres cannot interpenetrate, and in fact
strictly move away from each other. The formal proof and a more detailed
discussion will be given in Ref. \cite{Second_paper}.
} that we only need to consider first derivatives%
\footnote{The formal statement is that a packing is \emph{rigid if and only
if it is infinitesimally rigid}, see Refs. \cite{Connelly_disks,Second_paper}.
} \( \VV =\frac{d}{dt}\VdR (t) \), which can be thought of as ``velocities'',
and then simply move the spheres in the directions \( \VV =\left( \Vv _{1},\ldots ,\Vv _{2}\right)  \)
to obtain an unjamming motion \( \VdR (t)=\VV t \). That is, a sphere
packing that is not jammed can be unjammed by giving the spheres velocities
\( \VV  \) such that no two contacting spheres \( i \) and \( j \),
\( \left\Vert \Vr _{i}-\Vr _{j}\right\Vert =D \), have a relative
speed \( v_{i,j} \) toward each other%
\footnote{This is the third definition (definition \emph{c}) in section 2.1
of Ref. \cite{Connelly_Tensegrities}.
}:\begin{equation}
\label{relative_velocity}
v_{i,j}=\left( \Vv _{i}-\Vv _{j}\right) ^{T}\Vu _{i,j}\leq 0
\end{equation}
where \( \Vu _{ij}=\frac{\Vr _{j}-\Vr _{i}}{\left\Vert \Vr _{i}-\Vr _{j}\right\Vert } \)is
the unit vector connecting the two spheres%
\footnote{The sign notation may be a bit unorthodox but is taken from Ref. \cite{Vanderbei}.
}. Of course, some special and trivial cases like rigid body translations
(\( \VV =\textrm{constant} \)) need to be excluded since they do
not really change the configuration of the system. 

In this paper we will plot unjamming motions as ``velocity'' fields,
and occasionally supplement such illustrations with a sequence of
frames from \( t=0 \) to \( t=1 \) showing the unjamming process.
Note that the lengths of the vectors in the velocity fields have been
scaled to aid in better visualization%
\footnote{Since this paper deals only with small displacements in an approximate
linearized fashion, one should do a line-search in \( t \) along
\( \VR +\VdR (t) \) to check when the first collision occurs. The
\noun{VRML} animations shown at the webpage of Ref. \cite{webpage}
also depict arbitrarily scaled displacements for visualization purposes
so that collisions might happen before \( t=1 \).
}. For the sake of clear visualization, only two-dimensional examples
will be used. But all of the techniques described here are fully applicable
to three-dimensional packings as well. Interactive Virtual Reality
Modeling Language (\noun{VRML}) animations which are very useful
in getting an intuitive feeling for unjamming mechanisms in sphere
packings can be viewed from \noun{Windows} platforms on our webpage
(see Ref. \cite{webpage}).

\section{\label{Boundary_Conditions}Boundary Conditions}

As previously mentioned, the boundary conditions imposed on a given
packing are very important, especially in the case of strict jamming.
Here we consider two main types of packings, depending on the boundary
conditions.

\begin{description}
\item [\label{hard_walls}Hard-wall~boundaries]The packing \( P(\VR ) \)
is placed in an impenetrable concave%
\footnote{A container is concave if its boundary is concave at every circular
point. All convex polyhedral sets make a concave container, but a
spherical container is not concave. The details of the definition
and the reasons why the container needs to be concave to make some
of the theorems used here work are given in Ref. \cite{Connelly_disks}.
} \emph{hard-wall container} \( K \). Figure \ref{Honeycomb.HW.unjamming}
shows that the honeycomb lattice can be unjammed inside a certain
hard-wall container. Often we will make an effective container out
of \( N_{f} \) \emph{fixed spheres} whose positions cannot change.
This is because it is often hard to fit a packing into a simple container
such as a square box, while it is easy to surround it with other fixed
spheres, particularly if a periodic lattice is used to generate the
packing. Specifically, one can take a finite sub-packing of an infinite
periodic packing and freeze the rest of the spheres, thus effectively
making a container for the sub-packing. An example is depicted in
Fig. \ref{Honeycomb.pinned.unjamming}.

 
\item [Periodic~boundaries]Periodic boundary conditions are often used
to emulate infinite systems, and they fit the algorithmic framework
of this work very nicely. A periodic packing \( \widehat{P}(\VR ) \)
is generated by a replicating a finite \emph{generating packing} \( P(\widehat{\VR }) \)
on a lattice \( \MLambda =\left\{ \Vlambda _{1},\ldots ,\Vlambda _{d}\right\}  \),
where \( \Vlambda _{i} \) are linearly independent \emph{lattice
vectors} and \( d \) is the spatial dimensionality. So the positions
of the spheres are generated by,\[
\Vr _{\widehat{i}\left( \Vnc \right) }=\widehat{\Vr }_{i}+\MLambda \Vnc \textrm{ and }\Vnc \textrm{ is integer},\textrm{ }\Vnc \in Z^{d}\]
 where we think of \( \MLambda  \) as a matrix having the lattice
vectors as columns%
\footnote{The matrix \( \MLambda  \) has \( d^{2} \) elements.
} and \( \Vnc  \) is the number of replications of the unit cell along
each basis direction%
\footnote{This can also be viewed as a simple way of generating an infinite
packing, and one can analyze the resulting infinite packing, called
a \emph{cover of \( \Re ^{d} \)}, as discussed in Ref. \cite{Connelly_packings},
however, infinite packings are mathematically delicate and will be
discussed in Ref. \cite{Second_paper}.
}. The sphere \( \widehat{i}\left( \Vnc \right)  \) is the familiar
\emph{image sphere} of the \emph{original sphere} \( i=\widehat{i}\left( 0\right)  \),
and of course for the impenetrability condition only the nearest image
matters. For true periodic boundary conditions, we wrap the infinite
periodic packing \( \widehat{P}(\VR ) \) around a flat torus%
\footnote{We mean a topological torus, \emph{not} a geometrical one, i.e., a
torus where distances are still measured as in flat Eucledian space
but which has the topology of a curved torus.
}, i.e. we ask that whatever happens to a sphere \( i \) also happens
to all of the image spheres \( \widehat{i}\left( \Vnc \right)  \),
with the additional provision that the lattice may also change by
\( \MdLambda  \):\begin{equation}
\label{periodic_BCs}
\Delta \Vr _{\widehat{i}\left( \Vnc \right) }=\Delta \Vr _{i}+(\MdLambda )\Vnc 
\end{equation}
 
\end{description}

\subsection{Using Simple Lattices to Generate Packings}

Simple familiar lattices%
\footnote{We mean lattices with a basis, to be more precise.
} such as the triangular, honeycomb, Kagom\'{e} and square in two dimensions,
or the simple cubic (SC), body-centered cubic (BCC), face-centered
cubic (FCC) and hexagonal-close packed (HCP) in three dimensions,
can be used to create a (possibly large) packing taking a subsystem
of size \( \VNc  \) unit cells along each dimension from the infinite
lattice packing. The properties of the resulting system can be studied
with the tools developed here, provided that we restrict ourselves
to finite \( \VNc  \). Moreover, it is important to specify which
lattice vectors are to be used. We will usually take them be primitive
vectors, but sometimes it will be more convenient to use conventional
ones (especially for variations on the cubic lattice).

For hard-wall boundary conditions, we can take an infinite packing
generated by these simple lattices and then freeze all but the spheres
inside the window of \( \VNc  \) unit cells, thus effectively obtaining
a hard-wall container. Figure \ref{Honeycomb.pinned.unjamming} illustrates
an unjamming motion for the honeycomb lattice under these conditions.


For periodic boundary conditions, the generator \( P(\widehat{\VR }) \)
can itself be generated using a simple lattice%
\footnote{In this case the lattice \( \MLambda  \) is a \emph{sub-lattice}
of the underlying (primitive) lattice \( \widetilde{\MLambda } \),
i.e., \( \Vlambda _{\alpha }=\left[ \VNc \right] _{\alpha }\widetilde{\Vlambda }_{\alpha } \)
is an integer multiple of the vector of the underlying lattice \( \widetilde{\Vlambda }_{\alpha } \),
\( \alpha =1,\ldots ,d \).
}. This is not only a convenient way to generate simple finite periodic
packings, but it is in general what is meant when one asks, for example,
to analyze the jamming properties of the Kagom\'{e} lattice under
periodic or hard-wall boundary conditions. Figure \ref{Kagome.periodic.unjamming}
shows a periodic unjamming motion for the Kagom\'{e} lattice. Notice
though that the jamming properties one finds depend on how many neighboring
unit cells \( \VNc  \) are used as the ``base'' region (i.e., the
generating packing), and therefore, we will usually specify this number
explicitly. Some properties are independent of \( \VNc  \), and tailored
mathematical analysis can be used to show this \cite{Connelly_disks_II,Connelly_packings}.
For example, if we find a periodic unjamming motion for \( \VNc =\left( 1,2,\ldots \right)  \),
then we can be sure that for any \( \VNc  \) that is an integer multiple
of this window we can use the same unjamming motion. Correspondingly,
if we show that the system with \( \VNc =\left( 2\cdot 3\cdot 5,2\cdot 3,\ldots \right)  \)
is jammed, then so must be any system whose size can be factored into
the same prime numbers along each dimension. We will not consider
these issues in detail here, but rather focus on algorithmic approaches
tailored for finite and fixed systems (i.e., \( \VNc  \) is fixed
and finite), and postpone the rest of the discussion to Ref. \cite{Second_paper}. 


\section{\label{algorithm}Linear Programming Algorithm to Test for Jamming}

Given a sphere packing, we would often like to test whether it is
jammed according to each of the categories given above, and if it
is not, find one or several unjamming motions \( \VdR (t) \). We
now describe a simple algorithm to do this that is exact for \emph{gap-less
packings}, i.e., packings where neighboring spheres touch exactly,
and for which the definitions given earlier apply directly. However,
in practice, we would also like to be able to study \emph{packings
with small gaps}, such as produced by various heuristic compression
schemes like the Lubachevsky-Stillinger algorithm \cite{LS_algorithm}.
In this case the meaning of unjamming needs to be modified so as to
fit physical intuition. We do this in such a way as to also maintain
the applicability of our efficient randomized linear programming algorithm
using what Roux \cite{Roux} calls the \emph{approximation of small
displacements} (ASD).

\subsection{Approximation of Small Displacements}

As already explained, an unjamming motion for a sphere packing can
be obtained by giving the spheres suitable velocities, such that neighboring
spheres do not approach each other. Here we focus on the case when
\( \VdR (t)=\VV t+\mathcal{O}(t^{2}) \) are small finite displacements
from the current configuration. Therefore, we will drop the time designation
and just use \( \VdR  \) for the displacements from the current configuration
\( \VR  \) to the new configuration \( \widetilde{\VR }=\VR +\VdR  \). 

In this ASD approximation, we can linearize the impenetrability constraints
by expanding to first order in \( \VdR  \), \begin{equation}
\label{Impenetrability_NL}
\left\Vert \widetilde{\Vr }_{i}-\widetilde{\Vr }_{j}\right\Vert =\left\Vert (\Vr _{i}-\Vr _{j})+(\Vdr _{i}-\Vdr _{j})\right\Vert \geq D
\end{equation}
to get the condition for the existence of a \emph{feasible displacement}
\( \VdR  \),\begin{equation}
\label{Impenetrability_L}
(\Vdr _{i}-\Vdr _{j})^{T}\Vu _{i,j}\leq \Delta l_{i,j}\textrm{ for all }\left\{ i,j\right\} 
\end{equation}
 where \( \left\{ i,j\right\}  \) represents a \emph{potential contact}
between nearby spheres \( i \) and \( j \), \( \Delta l_{i,j}=\left\Vert \Vr _{i}-\Vr _{j}\right\Vert -D \)
is the \emph{interparticle gap}%
\footnote{Called \emph{interstice} in Ref. \cite{Roux}.
}, and \( \Vu _{ij}=\frac{\Vr _{j}-\Vr _{i}}{\left\Vert \Vr _{i}-\Vr _{j}\right\Vert } \)
is the unit vector along the direction of the contact \( \left\{ i,j\right\}  \).
For a gap-less packing, we have \( \Delta l=0 \) and the condition
(\ref{Impenetrability_L}) reduces to (\ref{relative_velocity}),
which as we explained along with the conditions \( \VdR \neq 0 \)
and \( \VdR \neq \textrm{const}. \) is an \emph{exact} condition
for the existence of an unjamming motion \( \VdR  \). For packings
with finite but small gaps though, condition (\ref{Impenetrability_L})
is only a first-order approximation. Notice that we only need to consider
potential contacts \( \left\{ i,j\right\}  \) between nearby, and
not all pairs of spheres%
\footnote{\label{Gap_tolerance}That is we only consider a contact if\[
\left\Vert \Vr _{i}-\Vr _{j}\right\Vert \leq \left( 1+\delta \right) D\]
where \( \delta  \) is some tolerance for how large we are willing
to allow the representative \( \left\Vert \Vdr \right\Vert  \) to
be. The larger this tolerance, the more possible particle contacts
we will add to our constraints, and thus the more computational effort
we need. Also, the ASD approximation becomes poorer as we allow larger
displacements. But choosing a very small tolerance makes it impossible
to treat systems with moderately large interparticle gaps (say of
the order of \( \delta =10\% \)).
}. The complicated issue of how well the ASD approximation works when
the gaps are not small enough is illustrated in Fig. \ref{LP_cones}.


By putting the \( \Vu _{ij} \)'s as columns in a matrix of dimension
\( \left[ Nd\times N_{e}\right]  \), where \( N_{e} \) is the number
of contacts in the contact network, we get the important \textbf{rigidity
matrix}%
\footnote{This is in fact the negative transpose of what is usually taken to
be the rigidity matrix, and is chosen to fit the notation in Ref.
\cite{Vanderbei}.
} of the packing \( \MA  \). This matrix is sparse and has two blocks
of \( d \) non-zero entries in the column corresponding to the particle
contact \( \left\{ i,j\right\}  \), namely, \( \Vu _{ij} \) in the
block row corresponding to particle \( i \) and \( -\Vu _{ij} \)
in the block row corresponding to particle \( j \). Represented schematically:

\[
\begin{array}{c}
\\
\MA 
\end{array}\begin{array}{c}
\\
=
\end{array}\begin{array}{cc}
 & \begin{array}{c}
\{i,j\}\\
\downarrow 
\end{array}\\
\begin{array}{c}
i\rightarrow \\
\\
j\rightarrow 
\end{array} & \left[ \begin{array}{c}
\vdots \\
\Vu _{ij}\\
\vdots \\
-\Vu _{ij}\\
\vdots 
\end{array}\right] 
\end{array}\]
For example, for the four-disk packing shown in Fig. \ref{LP_cones},
and with the numbering of the disks depicted in Fig. \ref{LP_cone_packing},
we have the following rigidity matrix:\[
\begin{array}{cc}
 & \begin{array}{ccccccc}
E_{12} &  & E_{13} &  & E_{14} & 
\end{array}\\
\begin{array}{c}
\MA \\

\end{array}\begin{array}{c}
=\\

\end{array}\begin{array}{c}
D_{1}\\
D_{2}\\
D_{3}\\
D_{4}
\end{array} & \left[ \begin{array}{cccc}
\Vu _{12} & \Vu _{13} & \Vu _{14} & \\
-\Vu _{12} &  &  & \\
 & -\Vu _{13} &  & \\
 &  & -\Vu _{14} & 
\end{array}\right] 
\end{array}\]


Using this matrix, we can rewrite the linearized impenetrability constraints
as a simple system of linear inequality constraints:\begin{equation}
\label{impenetrability}
\MA ^{T}\VdR \leq \Vdl 
\end{equation}

The set of contacts \( \left\{ i,j\right\}  \) that we include in
(\ref{Impenetrability_L}) form the \textbf{contact network} of the
packing, and they correspond to a subclass of the class of fascinating
objects called \emph{tensegrity frameworks}, namely \emph{strut frameworks}
(see Ref. \cite{Connelly_Tensegrities} for details, and also \cite{Tensorial_contact}
for a treatment of more general packings). Figure \ref{Contact_100}
shows a small random packing with relatively large gaps and the associated
contact network.


\subsubsection{Boundary conditions}

Handling different boundary conditions within the above formulation
is easy. For example, for usual periodic conditions, handling the
boundaries merely amounts to adding a few columns to the rigidity
matrix \( \MA  \) with \( \Vu _{i,\widehat{j}(\Vnc )}=\frac{\Vr _{\widehat{j}(\Vnc )}-\Vr _{i}}{\left\Vert \Vr _{\widehat{j}(\Vnc )}-\Vr _{i}\right\Vert } \)
for all images \( \widehat{j}(\Vnc ) \) which have contacts with
one of the original spheres \( i \). These columns correspond to
the periodic contacts wrapping the packing around the torus. 

For hard-wall boundaries, we would add a potential particle contact
to the contact network from each sphere close to a wall to the closest
point on the wall, and fix this endpoint. Such fixed points of contact
and fixed spheres%
\footnote{Such nodes are called \emph{fixed nodes} in tensegrity terminology.
} \( j \) are simply handled by transferring the corresponding term
\( \Vdr _{j}^{T}\Vu _{i,j} \) to the right-hand side of the constraints
in (\ref{impenetrability}).

\subsection{Finding Unjamming Motions}

We are now ready to explain how one can find unjamming motions for
a given packing, if such exist. But first, we need to refine our definition
of an unjamming motion to allow for the study of packings with finite
gaps.

\subsubsection{Unjamming Motions Revisited: \emph{Dealing with Finite Gaps}}

The problem with packings with small gaps is that according to our
previous definition of an unjamming motion in section \ref{Unjamming_motions},
such packings will never be jammed, since it will always be possible
to move some of the spheres at distances comparable to the sizes of
the interparticle gaps so that they lose contact with all or most
of their neighbors. Clearly we wish to only consider the possibility
of displacing some of the spheres such that \emph{considerable} gaps
appear, larger then some threshold value \( \Delta l_{\textrm{large}}\gg \overline{\Delta l} \),
where \( \overline{\Delta l} \) is a measure of the magnitude of
the interparticle gaps. Therefore, we have the modified definition:
An \textbf{unjamming} \textbf{motion} \( \VdR (t) \), \( t\in \left[ 0,1\right]  \),
is a \emph{continuous} displacement of the spheres from their current
position along the path \( \VR +\VdR (t) \), starting from the current
configuration, \( \VdR (0)=0 \), and observing all relevant constraints
along the way, such that some of the spheres lose contact%
\footnote{Alternatively, we could ask that some sphere displace by more than
\( \Delta r_{\textrm{large}}\gg \overline{\Delta l} \).
} with each other in the final configuration \( \VR +\VdR (1) \) by
more then a given \( \Delta l_{\textrm{large}} \). Again, we exclude
any trivial rigid-body type motions such as a uniform translation
of the spheres from consideration.

The problem of whether such an unjamming motion exists and how to
find one if it does is mathematically very interesting
if we take the case \( \Delta l_{\textrm{large}}\gg D \). But this
problem is extremely complex due to high non-linearity of the impenetrability
constraints and we will make no attempt to solve it. It also is not
clear that this would be of interest to physicists. Instead, we focus
our attention on the case when \( \Delta l_{\textrm{large}} \) is
small enough to apply with a reasonable degree of accuracy the approximation
of small displacements, but large enough compared to the interparticle
gaps so that the exact value is really irrelevant. 

Under the ASD, we need only worry about linear displacements from
the current configuration, \( \VdR (t)=\VV t \), and so we can focus
on \( \VdR \equiv \VdR (1) \). Thus, finding an unjamming motion%
\footnote{Since here we are really focusing only on \( \VdR \equiv \VdR (1) \),
we should change terminology and call \( \Delta r \) an \textbf{unjamming
displacement}. However, to emphasize the fact that there is a way
to continuously move the spheres to achieve this displacement, and
also that we can always scale \emph{down} the magnitude of an allowed
\( \VdR  \) arbitrarily, we continue to say unjamming motion. The
term ``displacement'' will be more appropriate in section \ref{Duality}.
} simply reduces to the problem of \emph{feasibility of a linear system
of inequalities,}

\begin{eqnarray}
 & \MA ^{T}\VdR \leq \Vdl  & \textrm{for impenetrability}\\
\exists \left\{ i,j\right\} \textrm{ such that } & \left| \left( \MA ^{T}\VdR \right) _{\left\{ i,j\right\} }\right| \geq \Delta l_{\textrm{large}}\gg \overline{\Delta l} & \textrm{for unjamming}\label{Feasibility_question} 
\end{eqnarray}
 where we can exclude trivial displacements such as uniform translations
by adding additional constraints (e.g., demanding that the centroid
remains fixed, \( \sum \Vdr _{i}=0 \)).

\subsubsection{Randomized Linear Programming (LP) Algorithm}

The question of whether a packing is jammed, i.e., whether the system
(\ref{Feasibility_question}) is feasible, can be answered rigorously%
\footnote{The gap-less case \( \Vdl =0 \), \( \Delta l_{\textrm{large}}\rightarrow 0^{+} \),
can be studied rigorously. When gaps are present, of course, the condition
\( \left| \left( \MA ^{T}\VdR \right) _{\left\{ i,j\right\} }\right| \gg \overline{\Delta l} \)
is mathematically ambiguous and also the ASD approximation becomes
inexact.
} by using standard linear programming techniques%
\footnote{\label{LP_feasibility_footnote}Solve the following linear program
aimed at maximizing the sum of the (positive) gap dilations \( \left( \Vdl -\MA ^{T}\VdR \right) _{i,j} \),\begin{eqnarray}
 & \min _{\VdR }\sum _{\left\{ i,j\right\} }(\MA ^{T}\VdR )_{i,j}=\min \left( \MA \mathbf{e}\right) ^{T}\VdR  & \\
\textrm{such that} & \MA ^{T}\VdR \leq \Vdl  & \label{LP_feasibility} 
\end{eqnarray}
where \( \mathbf{e} \) is the unit vector, and then look at the magnitudes
of the gap dilations (these may be unbounded of course) and decide
if they are large enough to consider the solution an unjamming motion.
Otherwise the packing is jammed. Notice that this will usually produce
a single unjamming motion, which we have found to be rather uninteresting
for lattice packings in the sense that it is extremely dependent upon
\( \VNc  \).
}. If a packing is jammed, then this is enough. But for packings which
are not jammed, it is really more useful to obtain a \emph{representative}
collection of unjamming motions. A random collection of such unjamming
motions is most interesting, and can be obtained easily by solving
several linear programs with a random cost vector.

We adopt such a \textbf{randomized LP algorithm} to testing unjamming
{[}i.e. studying (\ref{Feasibility_question}){]}, namely, we solve
several instances of the following LP in the \emph{displacement formulation}:\begin{eqnarray}
 & \max _{\VdR }\Vb ^{T}\VdR  & \textrm{for virtual work}\\
\textrm{such that} & \MA ^{T}\VdR \leq \Vdl  & \textrm{for impenetrability}\label{LP_displacement} \\
 & \left| \VdR \right| \leq \Delta R_{\textrm{max}} & \textrm{for boundedness}
\end{eqnarray}
for \emph{random loads}%
\footnote{The physical interpretation of \( \Vb  \) as an external load will
be elucidated in section \ref{Duality}.
} \( \Vb  \), where \( \Delta R_{\textrm{max}}\gg \overline{\Delta l} \)
is used to prevent unbounded solutions and thus improve numerical
behavior%
\footnote{\label{b_choice}Even for jammed packings, unless \( \Vb  \) is chosen
carefully, the solution of (\ref{LP_displacement}) will be unbounded
due to the existence of trivial motions such as uniform translations.
This will become clear once duality is discussed, but mathematically
\( \Vb  \) needs to be in the null-space of \( \MA  \), which usually
means it needs to have zero total sum and total torque (see chapter
15 in Ref. \cite{Vanderbei}). 
}. Trivial solutions, such as uniform translations of the packing \( \VdR =\textrm{const}. \)
for periodic boundary conditions, can be eliminated \emph{a posteriori},
for example by reducing \( \VdR  \) to zero mean displacement%
\footnote{The choice of \( \Vb  \) also affects the appearance of these trivial
solutions in the optimal solution, which is non unique, as explained
in footnote \ref{b_choice}.
}, or added as extra constraints in (\ref{LP_displacement}). We will
discuss numerical techniques to solve (\ref{LP_displacement}) shortly.

We then treat any solution \( \VdR  \) to (\ref{LP_displacement})
with components significantly larger then \( \overline{\Delta l} \)
as an unjamming motion. For each \( \Vb  \), if we fail to find an
unjamming motion, we apply \( -\Vb  \) as a loading also%
\footnote{\label{Polyhedrons_cones}The linearized impenetrability constraints
\( \MA ^{T}\VdR \leq \Vdl  \) define a polyhedral set \( \mathcal{P}_{\VdR } \)
of \emph{feasible displacements}. Every such polyhedron consists of
a finite piece \( \mathcal{P}_{\VdR }^{\textrm{hull}} \), the convex
hull of its extreme points, and possibly an unbounded piece \( \mathcal{C}_{\VdR } \),
a finitely generated \emph{polyhedral cone}. In some cases this cone
will be empty (i.e. \( \mathcal{C}_{\VdR }=\left\{ 0\right\}  \)),
but in others it will not, as can be seen in Fig. \ref{LP_cones}.
A mathematically very well defined formulation is to take any ray
in the cone \( \mathcal{C}_{\VdR } \) as an unjamming motion, and
exclude others, however, as Fig. \ref{LP_cones} shows, the elongated
corners of this polyhedron are in fact very likely to be unbounded
in the true non-linear feasible set of displacements, so we prefer
to take any ``long'' direction in \( \mathcal{P}_{\VdR } \) as an
unjamming motion.\\
\\
We note that the randomized LP algorithm proposed here strictly answers
the question of whether the polyhedral set of feasible displacements
contains an unbounded ray just by applying two (nonzero) loads \( \Vb  \)
and \( -\Vb  \). This is because an attempt to find such a ray will
be unsuccessful only if \( -\Vb \in \mathcal{C}_{\VdR }^{*} \), where
\( \mathcal{C}_{\VdR }^{*} \) is the conjugate cone of \( \mathcal{C}_{\VdR }^{*} \),
and in this case \( \Vb \notin \mathcal{C}_{\VdR }^{*} \), so that
using the load \( -\Vb  \) will find a ray if such a ray exists.
Also, we note that one cannot hope to fully characterize the cone
of first-order unjamming motions \( \mathcal{C}_{\VdR }^{*} \) (i.e.
find its convex hull of generating rays), as this is known to be an
NP complete problem related to the full enumeration of the vertices
of a polyhedron. Our randomized approach essentially finds a few sample
rays in \( \mathcal{C}_{\VdR }^{*} \).
}. In our tests we usually set \( \Delta R_{\textrm{max}}\sim 10D \)
and treated any displacement where some gap dilations \( \left( \MA ^{T}\VdR \right) _{i,j}\sim D \)
as unjamming motions, but as should be clear from the discussion in
footnote \ref{Polyhedrons_cones}, the method is not very sensitive
to the exact values. One can then save these individual unjamming
motions and visualize them, or try to combine several of these into
one ``most interesting'' unjamming motion%
\footnote{We believe motions in which as many of the spheres move as possible
are most useful since then one can see multiple unjamming ``mechanisms''
with one visualization. Therefore, we make a convex combination of
several unjamming motions \( \VdR (\Vb _{\alpha }) \) obtained from
several different random loads \( \Vb _{\alpha } \) to obtain one
such unjamming motion.
}, as we have done to obtain the figures in this paper.

We stress that despite its randomized character, this algorithm is
almost rigorous when used as a test of jamming, in the sense that
it is \emph{strictly rigorous} for gap-less packings and also for
packings with small gaps as explained in more detail in footnote \ref{Polyhedrons_cones}.

\section{Testing for Local, Collective and Strict Jamming: \emph{Periodic
Boundary Conditions}}

\subsection{Local Jamming}

Recall that the condition for a packing to be locally jammed is that
\emph{each particle be fixed by its neighbors}. This is easy to check.
Namely, each sphere has to have at least \( d+1 \) contacts with
neighboring spheres, not all in the same \( d \)-dimensional hemisphere.
This is easy to test in any dimension by solving a small linear program,
and in two and three dimensions one can use more elementary geometric
constructions. 

We prefer the LP approach because it is in the spirit of this work
and because of its dimensional independence, and so we present it
here. Take a given sphere \( i \) and its set of contacts \( \left\{ \Vu _{i*}\right\}  \),
and put these as rows in a matrix \( \MA ^{T}_{i} \). Then solve
the local portion of (\ref{LP_feasibility}) in footnote \ref{LP_feasibility_footnote}
(using the simplex algorithm):\begin{eqnarray}
 & \min _{\Vdr _{i}}(\MA _{i}\mathbf{e})^{T}\Vdr _{i} & \\
\textrm{such that} & \MA ^{T}_{i}\Vdr _{i}\leq \Vdl _{i,*} & \label{LP_local} 
\end{eqnarray}
which will have an unbounded solution if the sphere \( i \) is not
locally jammed, as illustrated in Fig. \ref{LP_cones}. 

Of course we can define higher orders of local jamming by asking that
\emph{each set of \( n \) spheres be fixed by its neighbors}, called
\emph{n-stability} in Ref. \cite{Connelly_packings}. However, for
\( n>1 \) it becomes combinatorially too difficult to check for this.
Computationally, we have found testing for local jamming using (\ref{LP_local})
to be quite efficient and simple.

\subsection{Collective Jamming}

The randomized LP algorithm was designed to test for collective jamming
in large packings, and in this case the linear program (\ref{LP_displacement})
that needs to be solved is very large and sparse. Notice that boundary
conditions are only involved when making the list of contacts in the
contact network and deciding if certain spheres or contact points
are fixed. In the case of periodic boundary conditions, we simply
add the usual contacts between original spheres near the boundary
of the unit cell and any nearby image spheres.

We have implemented an efficient numerical solution of (\ref{LP_displacement})
{[}and also (\ref{LP_force}){]}, using the primal-dual interior-point
algorithm in the \noun{LOQO} optimization library (see Ref. \cite{LP_libraries}).
Illustrations of results obtained using this implementation are given
throughout this paper. 

We would like to stress that primal-dual interior-point algorithms
are very well suited for problems of this type, and should also be
intuitive to physicists since in essence they solve a sequence of
easier problems in which the perfectly rigid inter-sphere contacts
are replaced by stiff (but still deformable) nonlinear (logarithmic)
springs, carefully numerically taking the limit of infinitely stiff
springs. Physicists have often used similar heuristically designed
schemes and hand-tuned them, and even suggested that standard optimization
algorithms are not practical (for example, in Ref. \cite{Roux}).
We would like to dispel such beliefs and stress the importance of
using robust and highly efficient software developed by applied mathematicians
around the world, such as Ref. \cite{LP_libraries}, becoming increasingly
more available. Not only are the algorithms implemented theoretically
well-analyzed, but they are tested on a variety of cases and often
contain several alternative implementations of computation-intensive
sections targeting different types of problems. Choice of the correct
algorithm and the details are often complex, but well worth the effort.

Nonetheless, for three-dimensional problems the available implementations
of interior-point algorithms based on direct linear solvers are too
memory demanding and inefficient. Tuned implementations based on conjugate-gradient
iterative solvers are needed. We plan to develop efficient parallel
algorithms suited for these types of problems and make them publicly
available in the very near future.

\subsection{\label{Strict_Jamming}Strict Jamming}

To extend the notion of collective jamming to strict jamming we introduced
deformations of the boundary. In the case of periodic packings, \emph{the
lattice \( \MLambda  \) is the boundary}. Therefore, the only difference
with collective jamming is that we will now allow the lattice to change
while the spheres move, i.e., \( \MdLambda \neq 0 \) in (\ref{periodic_BCs}).
The \textbf{lattice deformations} \( \MdLambda  \) will also become
unknowns in (\ref{LP_displacement}), but since these too enter linearly
in (\ref{periodic_BCs}), we still get a linear program, only with
coefficient matrix \( \MA  \) augmented with new (denser) rows in
the columns corresponding to contacts across the periodic boundary.
The actual implementation now requires more care and bookkeeping,
but the conceptual changes should be clear, and the randomized LP
algorithm remains applicable.

Obviously, we cannot allow the volume of the unit cell to enlarge,
since the unit cell is in a sense the container holding the packing
together. Therefore, we only consider \emph{volume-non-increasing
continuous lattice deformations} \( \MdLambda (t) \): \emph{\begin{equation}
\label{volume_preservation_NL}
\det \left[ \widetilde{\MLambda }=\MLambda +\MdLambda (t)\right] \leq \det \MLambda \textrm{ for }t>0
\end{equation}
}Through a relatively simple mathematical analysis to be presented
in Ref. \cite{Second_paper}, it can be shown that the principles
that applied to unjamming motions of the spheres \( \VdR  \) still
remain valid even when we extend the notion of an unjamming motion
to include a deforming lattice%
\footnote{That is, we now think of \( \left[ \VdR (t),\MdLambda (t)\right]  \)
as an unjamming motion.
}. That is, we can still only focus on linear motions \( \MdLambda (t)=\mathbf{W}t \),
\( \mathbf{W}=\textrm{const}. \) and the final small deformations
\( \MdLambda =\MdLambda (1) \), and need to consider only first-order
linearizations of the impenetrability (\ref{Impenetrability_NL})
and non-expansion (\ref{volume_preservation_NL}) nonlinear constraints%
\footnote{The condition (\ref{symmetric_deformation}) needs to hold also.
}. 

The linearized version of (\ref{volume_preservation_NL}) is:\begin{equation}
\label{volume_preservation_L}
\textrm{Tr}[(\MdLambda )\MLambda ^{-1}]\leq 0
\end{equation}
and this is just one extra linear constraint to be added to the linear
program (\ref{LP_displacement}). An extra condition which needs to
be added is that \emph{}\( (\MdLambda )\MLambda ^{-1} \) be symmetric,
which is also an added linear constraint,\begin{equation}
\label{symmetric_deformation}
(\MdLambda )\MLambda ^{-1}=\Mepsilon \textrm{ where }\Mepsilon =\Mepsilon ^{T}
\end{equation}
where we add \( \Mepsilon  \) as an unknown in the randomized LP
algorithm. This condition in fact does nothing more then eliminate
trivial rotations%
\footnote{Rotations of the lattice turn out to increase the unit-cell volume
at the second-order derivative, even though they are volume-preserving
up to first order.
} of the lattice%
\footnote{But one should still deal with trivial motions \emph{a posteriori}
with some care in certain pathological cases.
}.

The motivation for the category of strict jamming and its above interpretation
in the periodic case should be clear: Changing the lattice in a volume
non-increasing way models \emph{macroscopic non-tensile strain} and
is therefore of great relevance to studying the macroscopic mechanical
properties of random packings (see Ref. \cite{Cones_strain}). In
fact, \( \Mepsilon \textrm{ }=(\MdLambda )\MLambda ^{-1} \) can be
interpreted as the ``macroscopic'' \emph{strain-tensor}, which explains
why it is symmetric and also trace-free for shear deformations. We
also again point out that strict jamming is (significantly) stronger
then collective jamming for periodic boundary conditions, particularly
in two-dimensional packings%
\footnote{This point will be elaborated in Ref. \cite{Second_paper}.
}. This point is illustrated in Fig. \ref{Connelly.3.strict.unjamming},
which shows an unjamming motion involving a deformation of the lattice,
even though this lattice packing is collectively jammed.


\subsection{Shrink-And-Bump Heuristic}

The following heuristic test for collective jamming has been suggested
in Ref. \cite{LS_algorithm}: Shrink the particles by a small amount
\( \delta  \) and then start the Lubachevsky-Stillinger molecular
dynamics algorithm with random velocities, and see if the system gets
unjammed. One would also slowly enlarge the particles back to their
original size while they bump around, so as to allow finite termination
of this test (within numerical accuracies). We call this the \emph{shrink-and-bump
heuristic.} The idea is that the vector of velocities takes on random
values in velocity space and if there is a direction of unjamming,
it will be found with a high probability and the system will unjam%
\footnote{The theory presented here suggests that if a packing is indeed not
collectively jammed and has a relatively large cone of unjamming motions
(see footnote \ref{Polyhedrons_cones}), it can be unjammed using
this type of heuristic with high probability. However, notice that
this cone can in principle be very small, so finding a ray in it may
be a low-probability occurrence. For packings with finite gaps, though,
the heuristic incorporates nonlinear effects, which is an advantage.
}. Animations of this process can be found at Ref. \cite{webpage}.

This kind of heuristic has the advantage of being very simple and
thus easy to implement and use, and it is also very efficient. The
real problem is not so much its indeterminacy, but its strong dependence
on the exact value of \( \delta  \). For example, animations showing
how the Kagom\'{e} lattice inside a container made of fixed spheres
(as in Fig. \ref{Honeycomb.pinned.unjamming}) can be unjammed with
a large-enough \( \delta  \), even though it is actually collectively
jammed under these boundary conditions, can be found at Ref. \cite{webpage}.
In fact, many jammed \emph{large} packings will appear unstable under
this kind of test, as motivated with the notion of \emph{uniform stability},
defined in Ref. \cite{Connelly_packings} and elaborated on in Ref.
\cite{Second_paper}.

\section{Additional Applications}

\subsection{Compressing Packings using Linear Programming}

In this work we emphasize the utility of the randomized linear programming
algorithm as a testing tool for jamming, and also for finding representative
unjamming motions. The unjamming motions one finds can be used inside
compression algorithms. For example, the Lubachevsky-Stillinger algorithm
may sometimes get stuck in a particular configuration even though
the configuration is not collectively or strictly jammed (particularly
in two dimensions, as explained later). The unjamming motion obtained
from the linear programming algorithm can then be used to continue
the compression%
\footnote{For example, one can displace the spheres by \( \VdR  \) and restart
the compression with random velocities or use initial velocities along
\( \VdR  \) to unjam the packing and continue the simulation.
}. 

Even more can be done with linear programming in this direction. For
example, one can ask the question of whether there is an unjamming
motion \( \VdR  \) in which all sphere contacts are lost%
\footnote{All contacts will be lost if \( \MA ^{T}\VdR \leq -\alpha D \), where
\( \alpha >0 \).
}. This can be done for example by solving the LP,\begin{eqnarray}
 & \max _{\VdR ,\varepsilon }\, \alpha  & \\
\textrm{such that} & \MA ^{T}\VdR \leq -\alpha D & \label{LP_compression} \\
 & 0\leq \alpha \leq 1 & \textrm{for boundedness}
\end{eqnarray}
By displacing the spheres by such a \( \VdR  \) we would create a
``cushion'' of free space around each sphere, so that we can actually
increase the radius of the spheres%
\footnote{The common radius could increase by at least \( \alpha D \).
} and thus increase the density, or equivalently, the packing fraction
\( \varphi  \). We believe these kinds of approaches to be too inefficient,
since solving a large-scale linear program is too expensive to be
used iteratively. This is reminiscent of the high-quality but rather
expensive compression algorithm of Zinchenko (see Ref. \cite{Zinchenko}). 

Our future work will focus on using mathematical programming algorithms
and rigidity theory to design high-quality algorithms for design of
packings with target properties, using systems of stiff nonlinear
springs as an intermediary.

\subsection{\label{Duality}Kinematic/Static Duality}

The subject of kinematic/static duality and its physical meaning and
implications are discussed at length in Ref. \cite{Roux}, and elsewhere
in various degrees of relevance and different perpectives \cite{Connelly_Tensegrities,Connelly_disks,Moukarzel_isostatic,Rigidity_theory,Torquato_jammed}.
Here we only comment on it because of its relevance to the randomized
LP algorithm for testing jamming, and leave further discussion of
this important subject to Ref. \cite{Second_paper}. 

The dual%
\footnote{Excluding the additional practical safeguard constraint \( \VdR \leq \VdR _{\textrm{max}} \),
which is added to avoid unbounded trivial or unjamming motions.
} of the displacement formulation LP (\ref{LP_displacement}) also
has a very physical interpretation and it gives us the \emph{interparticle
repulsive}%
\footnote{We choose a negative sign for repulsive forces here in agreement with
mathematical literature \cite{Connelly_Tensegrities}.
} \emph{forces} \( f \) as dual variables, and we call it the \textbf{\emph{force
formulation}} LP:\begin{eqnarray}
 & \max _{\Vf }(\Vdl )^{T}\Vf  & \textrm{for virtual work}\\
\textrm{such that} & \MA \Vf =\Vb  & \textrm{for equilibrium}\label{LP_force} \\
 & \Vf \leq 0 & \textrm{for repulsion only}
\end{eqnarray}
The physical interpretation of the objective function in both the
displacement (\ref{LP_displacement}) and force formulations (\ref{LP_force})
is that of (virtual) mechanical work done by the external force load
\( \Vb  \) applied to the spheres. These two LP's are of great importance
in studying the \emph{stress-strain} behavior of granular materials,
as explained in Ref. \cite{Roux}, and since they are equivalent to
each other, we can call them the \emph{ASD stress-strain LP}. 

We wish to emphasize that by using primal-dual interior point algorithms
we automatically get both forces and displacements using the same
implementation%
\footnote{For example, both \noun{LOQO} and \noun{PCx} (\cite{LP_libraries})
return both primal and dual solutions to the user.
}. We have emphasized the displacement formulation (\ref{LP_displacement})
simply because we based our discussion of jamming on a kinematic perspective,
but a parallel static interpretation can easily be given. For example,
a random \( \Vb  \) used in the randomized LP algorithm that finds
an unbounded unjamming motion physically corresponds to a load that
the packing cannot support, i.e. the force formulation LP is (dual)
infeasible, implying that the displacement formulation LP is (primal)
unbounded%
\footnote{Interior-point algorithms deal better with unboundedness or infeasibility
in this context then the simplex algorithm.
}. The meaning of collective jamming within the ASD in the presence
of small gaps from a static standpoint now becomes clear: A collectively
jammed packing can resist (support) any force loading by (as) small
(as possible) rearrangements of the spheres, in which some of the
potential contacts are open and others closed, depending on the loading
and the interparticle gaps. 

In general the stress-strain LP will be highly degenerate and its
primal and/or dual solution not unique. However, as Roux points out,
the existence of small gaps in random packings is very important in
this context. Namely, if \( \Vdl  \) is random and nonzero (however
small), and \( \Vb  \) is also random, theorems on the generic character
of linear programs (see the references in Ref. \cite{Vanderbei})
can be invoked to guarantee that both the primal and dual solutions
will be non-degenerate. A non-degenerate solution to (\ref{LP_force})
corresponds to an isostatic force-carrying contact network, a fact
noted and explained in a great many ways by various researchers in
the field of granular materials \cite{Roux,Moukarzel_isostatic,Rigidity_theory}.
We just mention these points here in order to stimulate interest among
the physical community in the very relevant results to be found in
the mathematical programming literature.

\section{Results}

We have applied the randomized LP algorithm to test for the different
jamming categories in practice. The primary aim of this work is not
to give exhaustive results, but rather to introduce a conceptual framework
and some algorithms. Nonetheless, in this section we present some
sample relevant results for both ordered and disordered periodic packings.

\subsection{Periodic Lattice Packings}

Table 1 in Ref. \cite{Torquato_jammed} gives a classification of
some common simple lattice packings into jamming categories for hard-wall
boundary conditions. Table \ref{lattices_table} reproduces this for
periodic boundary conditions. As we explained in section \ref{Boundary_Conditions},
the results in principle will depend on the number of unit cells \( N_{c} \)
chosen as the original packing, and also on the unit cell chosen,
so the terminology ``lattice X is Y jammed'' is used loosely here.

\begin{table*}[!h]
{\centering \begin{tabular}{|c|c|c|c|c|c|c|c|c|}
\hline 
\emph{Lattice}&
\( \varphi  \)&
\emph{L}&
\( Z \)&
\emph{C}&
 \emph{}\( \VNc  \)&
S&
\( \VNc  \)&
 \( N_{s} \)\\
\hline
\hline 
Honeycomb&
\( 0.605 \)&
Y&
3&
N&
\( \left( 2,1\right)  \), \( \left( 1,2\right)  \)&
N&
\( \left( 1,1\right)  \)&
2\\
\hline 
Kagom\'{e}&
\( 0.680 \)&
Y&
6&
N&
\( \left( 1,1\right)  \)&
N&
\( \left( 1,1\right)  \)&
3\\
\hline 
Square&
\( 0.785 \)&
Y&
4&
N&
\( \left( 2,1\right)  \)&
N&
\( \left( 1,1\right)  \)&
1\\
\hline 
Triangular&
\( 0.907 \)&
Y&
6&
Y&
&
Y&
&
1\\
\hline 
Diamond&
\( 0.340 \)&
Y&
4&
N&
\( \left( 1,1,2\right)  \)&
N&
\( \left( 1,1,1\right)  \)&
2\\
\hline 
SC&
\( 0.524 \)&
Y&
6&
N&
\( \left( 1,1,2\right)  \)&
N&
\( \left( 1,1,1\right)  \)&
1\\
\hline 
BCC&
\( 0.680 \)&
Y&
8&
N&
\( \left( 1,1,2\right)  \)&
N&
\( \left( 1,1,1\right)  \)&
1\\
\hline 
FCC&
\( 0.741 \)&
Y&
12&
Y&
&
Y&
&
1\\
\hline 
HCP&
\( 0.741 \)&
Y&
12&
Y&
&
Y&
&
2\\
\hline
\end{tabular}\par}

\caption{\label{lattices_table}\emph{Classification of some simple lattices
into jamming categories}. We give the packing (i.e., covering) fraction
\protect\( \varphi \protect \) (to three decimal places), the coordination
number \protect\( Z\protect \), and the number of disks/spheres \protect\( N_{s}\protect \)
per unit cell, an assessment of whether the lattice is locally (L),
collectively (C) or strictly (S) jammed (Y is jammed, N is not jammed),
and the ``smallest'' number of unit cells \protect\( \VNc \protect \)
on which an unjamming motion exists (illustrated at Ref. \cite{webpage}).}
\end{table*}

It turns out that in the cases given in Table \ref{lattices_table},
the packings we have classified as not collectively or strictly jammed
will not be so for any large \( \VNc  \). Here we give the smallest
\( \VNc  \) for which we have found unjamming motions, and illustrate
some of these in Figs. \ref{Kagome_Honeycomb.collective.unjamming}
and \ref{Honeycomb.strict.unjamming}. 

Moreover, the packings classified as jammed, in this case being the
maximal density packings in two dimensions (triangular) and three
dimensions (FCP and HCP) will be so for any finite \( \VNc  \). We
leave justification and further discussion of this to Ref. \cite{Second_paper}.
Here we just point out for the curious that the triangular lattice
is not the only strictly jammed ordered disk packing%
\footnote{Conditions for strict jamming and other possibilities for strictly
jammed two-dimensional periodic packings will be discussed in Ref.
\cite{Second_paper}.
}; two other examples are shown in Fig. \ref{Kagome_reinforced} (see
Ref. \cite{Stillinger_lattices}).




\subsection{Periodic Random Packings}

We also tested a sample of periodic random packings in two and three
dimensions produced via the Lubachevsky-Stillinger compression algorithm
\cite{LS_algorithm} at different compression rates. This algorithm
often tends to produce a certain number of rattlers, i.e., spheres
which are not locally jammed, which we \emph{remove}%
\footnote{In the actual LP implementation, we freeze and ignore such particles.
} before testing for jamming%
\footnote{Notice that checking each sphere for local jamming using (\ref{LP_local})
only once is not enough under this removal scheme. Specifically, once a
rattling sphere is removed, this may remove some contacts from the packing
and can make other spheres not locally jammed. Therefore, 
neighbors of rattlers are recycled on a stack of spheres
to be checked for local jamming. We have observed that often, particularly
in two-dimensional LS packings, \emph{all disks can eventually be removed} on the
basis of just the local jamming test starting with only a few percent rattlers. 
}. We would like to stress that these are not comprehensive tests,
but they do illustrate some essential points, and so instead of giving
tables with statistics, we give some representative illustrations.
The tolerances for the interparticle gaps (see footnote \ref{Gap_tolerance})
used were in the range \( \delta \in \left[ 0.25,0.50\right]  \),
and, as explained earlier, the results in some cases depend on this
chosen tolerance, but not strongly.

\emph{All random disk (i.e., two-dimensional) packings we tested were}
\textbf{\emph{not}} \emph{strictly jammed}. At the typical LS end
states of roughly \( \varphi \approx 0.82 \), we generally found
that the packings were collectively jammed (with some exceptions such
as the packing shown in Fig. \ref{LS.dilute.collective.unjamming}),
although not strictly jammed, as with the packing depicted in Fig.
\ref{LS.dilute.strict.unjamming}. However, even at very high densities
(\( \varphi \approx 0.89 \)) the packings were only collectively
jammed, as illustrated in Fig. \ref{LS.dense.strict.unjamming}. Note
that quite different properties were observed for the three-dimensional
packings: \emph{All random sphere (i.e., three-dimensional) packings
we tested were strictly (and thus collectively) jammed}.




\section{Discussion}

Our results have important implications for the classification of
random disk and sphere packings and suggest a number of interesting
avenues of inquiry for future investigations. Random disk packings
are less well-understood then sphere packings. The tendency of disk
packings to {}``crystalize'' (to form ordered, locally dense domains)
at sufficiently high densities is well established. For example, Quickenden
and Tan \cite{RCP_disks} experimentally estimated the packing fraction
of the {}``random close packed'' (RCP) state to be \( \varphi \approx 0.83 \)
and found that the packing fraction could be further increased until
the maximum value of \( \varphi =0.906 \) was achieved for the triangular
lattice packing. By contrast, typical random sphere packings at \( \varphi  \)
in the range \( 0.63-0.66 \) cannot be further densified.

Our recent understanding of the ill-defined nature of random close
packing and of jamming categories raises serious questions about previous
two-dimensional studies, particularly the stability of such packings.
Our present study suggests that random disk packings are not strictly
jammed at \( \varphi \approx 0.83 \); at best they may be collectively
jammed. Of course, the old concept of the RCP state
was invalid in that it did
not account for the jamming category of the packing. Previous attempts
to estimate the packing fraction of the {}``random loose'' state
\cite{Local_stability_2D} are even more problematic, given that this
term is even less well-defined then the RCP state. The best way to
categorize random disk packings is to determine the maximally random
jammed (MRJ) state (see Ref. \cite{Torquato_MRJ}) for each of the
three jamming categories (local, collective and strict). Such investigations
have been initiated \cite{Anu_order_metrics} and
will be carried out in the future.

The identification of the MRJ state for strictly jammed \emph{disk}
packings is an intriguing open problem. On the one hand, we have shown
that random packings exist with densities in the vicinity of the maximum
possible value (\( \varphi =\pi / 2\sqrt{3} \)) that are not
strictly jammed, and on the other hand, there is a conjectured achievable
lower bound \( \varphi \geq \sqrt{3}\pi / 8 \) corresponding
to the ``reinforced'' Kagom\'{e} lattice (see Fig. \ref{Kagome_reinforced}).
For random sphere packings, an initial study undertaken in Ref. \cite{Anu_order_metrics},
using the LP algorithm described in this work, found that maximally
disordered random packings around \( \varphi \approx 0.63 \) were
strictly jammed. This suggests a close relation between the conventionally
accepted RCP packing fraction and the packing fraction of the 
MRJ state for strictly jammed packings. However,
it has been shown \cite{Anu_order_metrics} that at a fixed packing 
fraction \( \varphi \approx 0.63 \) the variation in the order can be substantial and 
hence packing fraction alone cannot completely characterize a random packing.
The conventionally accepted RCP packing fraction in two dimensions may be 
approximately close in value
to the packing fraction of the MRJ state for \emph{collectively} jammed packings.
Much less obvious is what is the MRJ state for collectively jammed sphere
packings. Finally, a completely unexplored question concerns the
identification of the MRJ state for locally jammed disk and sphere
packings.

\section{Conclusions}

In this work we have proposed, implemented, and tested a practical
algorithm for verifying jamming categories in finite sphere packings
based on linear programming, demonstrated its simplicity and utility,
and presented some representative results for ordered lattices and
random packings. Interestingly, the random packings that we tested
were strictly jammed in three dimensions, but \emph{not} in two dimensions.
Future applications of the randomized linear-programming algorithm
are to be expected. We will further present and explore the theoretical
connections between rigidity and jamming, kinematic and static rigidity,
rigidity and energy, rigidity and stability, and finite, periodic
and infinite packings in Ref. \cite{Second_paper}, and work is already
under way to provide highly efficient implementations of various optimization
algorithms for linear and nonlinear programming on large-scale (contact)
networks.

\begin{acknowledgments}
The authors would like to thank Robert Vanderbei for providing us
with the \noun{LOQO} optimization library.
\end{acknowledgments}

\bibliographystyle{revtex}

\newpage
\section{Figure Captions}

\begin{figure*}[!h]
\caption{\emph{\label{Honeycomb.HW.unjamming}Unjamming the honeycomb lattice}
inside a hard-wall simple box container. The arrows in the figures
given here show the direction of motion of the spheres \protect\( \VV \protect \)
in the linear unjamming motion, scaled by some arbitrary constant
to enhance the figure. \protect\( \VNc =\left( 3,2\right) \protect \)
unit cells make this small packing.}
\end{figure*}

\begin{figure*}[!h]
\caption{\label{Honeycomb.pinned.unjamming}\emph{Unjamming the honeycomb
lattice}. A sub-packing of size \protect\( \VNc =\left( 3,3\right) \protect \)
of an infinite honeycomb lattice packing is pinned by freezing all
neighboring image disks. A representative unjamming motion is shown
as a sequence of several frames between times \protect\( t=0\protect \)
and \protect\( t=1\protect \). The unshaded disks represent the particles
in the generating packing \protect\( P(\widehat{\VR })\protect \),
while the shaded ones are image disks that touch one of the original
disks.}
\end{figure*}

\begin{figure*}[!h]
\caption{\label{Kagome.periodic.unjamming}\emph{Unjamming the Kagom}\'{e}
\emph{lattice}. Periodic boundary conditions are used with \protect\( \VNc =\left( 2,2\right) \protect \).}
\end{figure*}

\begin{figure*}[!h]
\caption{\label{LP_cones}\emph{Feasible displacements polyhedron}. The figures
show three stationary (dark gray) disks surrounding a mobile disk
(light gray). For each of the three stationary disks, we have a nonlinear
impenetrability constraint that excludes the mobile disk from a disk
of radius \protect\( D\protect \) surrounding each stationary disk
(dark circles). Also shown are the linearized versions of these constraints
(dark lines), which are simply tangents to the circles at the point
of closest approach, as well as the region of feasible displacements
bounded by these lines (shaded gray). }

{\raggedright This region is a polyhedral set, and in the left figure
it is bounded, meaning that within the ASD the mobile disk is locally
jammed (trapped) by its three neighbors, while on the left it is unbounded,
showing the cone of locally unjamming motions (escape routes). Notice
that with the true nonlinear constraints, the mobile disk can escape
the cage of neighbors in both cases, showing that the ASD is not exact.
However, it should also be clear that this is because we have relatively
large interparticle gaps here.\par}
\end{figure*}

\begin{figure*}[!h]
\caption{\label{LP_cone_packing}The packing from Fig. \ref{LP_cones} shown
again with a numbering of the disks. \protect\( D_{i}\protect \)
denotes particle \protect\( i\protect \) and \protect\( E_{ij}\protect \)
denotes the contact between the \protect\( i\protect \)th and \protect\( j\protect \)th
particles, i.e., the contact \protect\( \left\{ i,j\right\} \protect \).}
\end{figure*}

\begin{figure*}[!h]
\caption{\emph{\label{Contact_100}Contact network} of a random packing of
100 disks with periodic boundary conditions and \protect\( \delta =0.5\protect \).
Periodic contacts with neighboring image spheres are also shown. All
disks are locally jammed within the rather large gap tolerance employed.}
\end{figure*}

\begin{figure*}[!h]
\caption{\emph{\label{Connelly.3.strict.unjamming}Example of a lattice deformation}.
The above periodic packing (packing 3 in Ref. \cite{Connelly_disks_II})
is collectively jammed, but not strictly jammed. It can be continuously
sheared toward the triangular lattice by deforming the lattice in
a volume-preserving manner, as shown here.}
\end{figure*}

\begin{figure}[!h]
\caption{\emph{\label{Kagome_Honeycomb.collective.unjamming}Simple collective
mechanisms in the Kagom}\'{e} \emph{and honeycomb lattices}, respectively.
These lattices are not collectively jammed with periodic boundary
conditions, as the sample unjamming motions periodic with \protect\( \VNc =\left( 1,1\right) \protect \)
for Kagom\'{e} (left) and \protect\( \VNc =\left( 1,2\right) \protect \)
for honeycomb (right) shown here illustrate.}
\end{figure}

\begin{figure*}[!h]
\caption{\emph{\label{Honeycomb.strict.unjamming}Shearing the honeycomb lattice}.
The honeycomb lattice is not strictly (or collectively) jammed, and
an example of a lattice deformation with \protect\( \VNc =\left( 1,1\right) \protect \)
is shown, replicated on several unit cells to illustrate the shear
character of the strain \protect\( \Mepsilon =(\MdLambda )\MLambda ^{-1}\protect \)
{[}c.f. (\ref{symmetric_deformation}){]}. Note that only three (original)
spheres are involved in the actual calculation of this unjamming motion,
the rest are image spheres.}
\end{figure*}

\begin{figure*}[!h]
\caption{\emph{\label{Kagome_reinforced}Examples of strictly jammed lattices}
in two dimensions. The \protect\( 6/7\protect \)th lattice (packing
number 2 in Ref. \cite{Connelly_disks_II} and the last packing in Ref. 
\cite{Stillinger_lattices}), left, is obtained by
removing every \protect\( 7\protect \)th disk from the triangular
lattice. The reinforced Kagom\'{e} lattice, right, is obtained by
adding an extra ``row'' and ``column'' of disks to the Kagom\'{e}
lattice and thus has the same density in the thermodynamic limit,
namely, it has every \protect\( 4\protect \)th disk removed from
the triangular packing (see also Ref. \cite{Stillinger_lattices}).}
\end{figure*}

\begin{figure*}[!h]
\caption{\label{LS.dilute.collective.unjamming}A 
random packing (\protect\( \varphi =0.82\protect \)) of 1000 disks 
that is \emph{not collectively jammed},
and a representative periodic unjamming motion.}
\end{figure*}

\begin{figure*}[!h]
\caption{\label{LS.dilute.strict.unjamming}A random packing (\protect\( \varphi =0.83\protect \))
of 1000 disks that is \emph{collectively jammed but not strictly jammed},
and a representative unjamming motion. Though it is hard to see from
this figure, this is indeed a shearing motion that induces an unjamming
mechanisms. A more insightful animation can be found at the webpage
\cite{webpage}.}
\end{figure*}

\begin{figure*}[!h]
\caption{\label{LS.dense.strict.unjamming}A \emph{dense} (\protect\( \varphi =0.89\protect \))
random packing of 1000 disks that is \emph{collectively
jammed but not strictly jammed}, and a representative unjamming motion.
One can see the grains gliding over each grain boundary due to the
shear, bringing this packing closer to a triangular lattice.}
\end{figure*}

\newpage
\section{Figures}
\setcounter{figure}{0}

\newpage
\begin{figure*}[!h]
   {\centering \resizebox*{1\columnwidth}{!}{\includegraphics{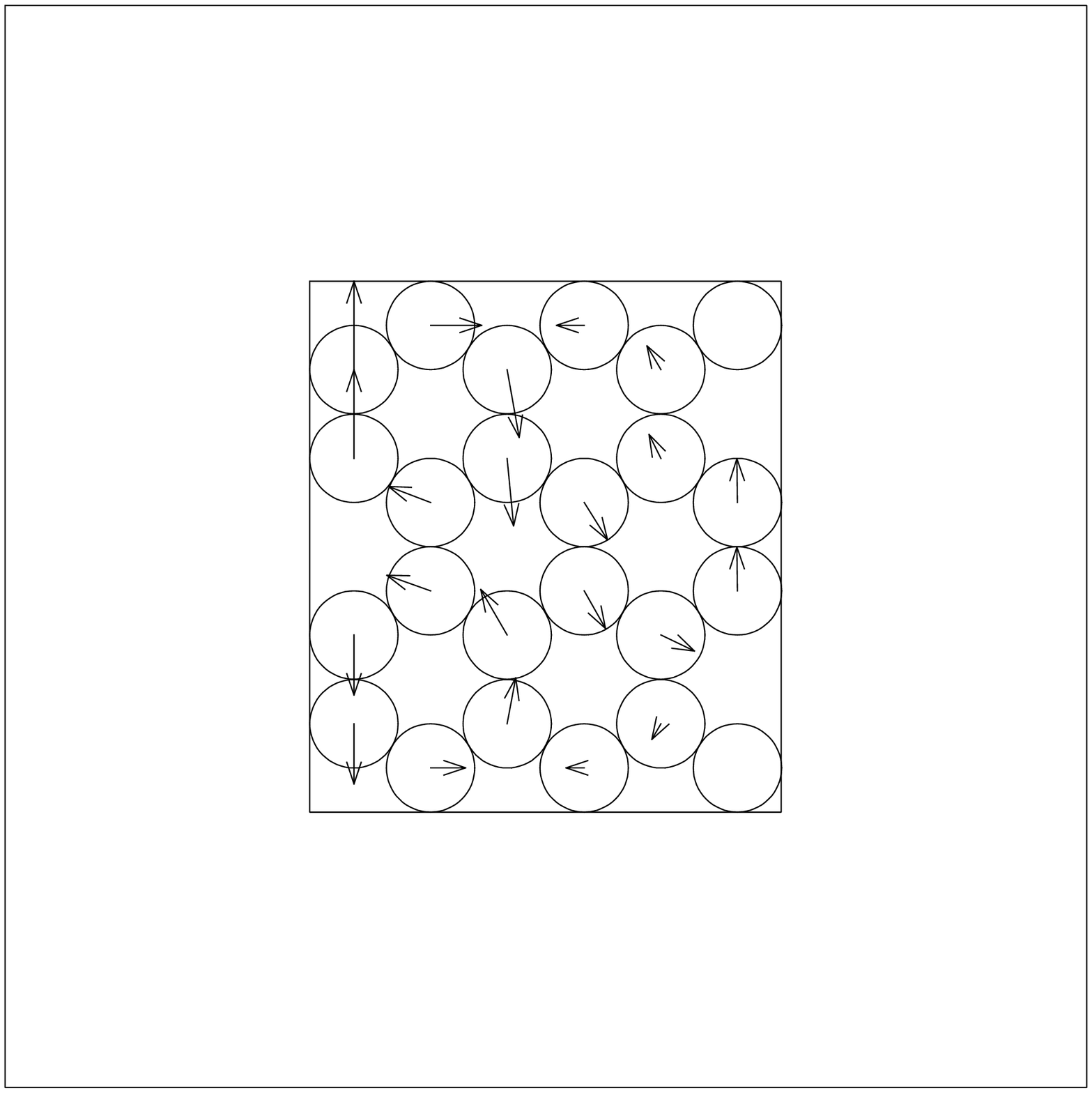}} \par}
   \caption{Donev \emph{et al.}}
\end{figure*}   

\newpage
\begin{figure*}[!h]
   {\centering \resizebox*{1\textwidth}{!}{\includegraphics{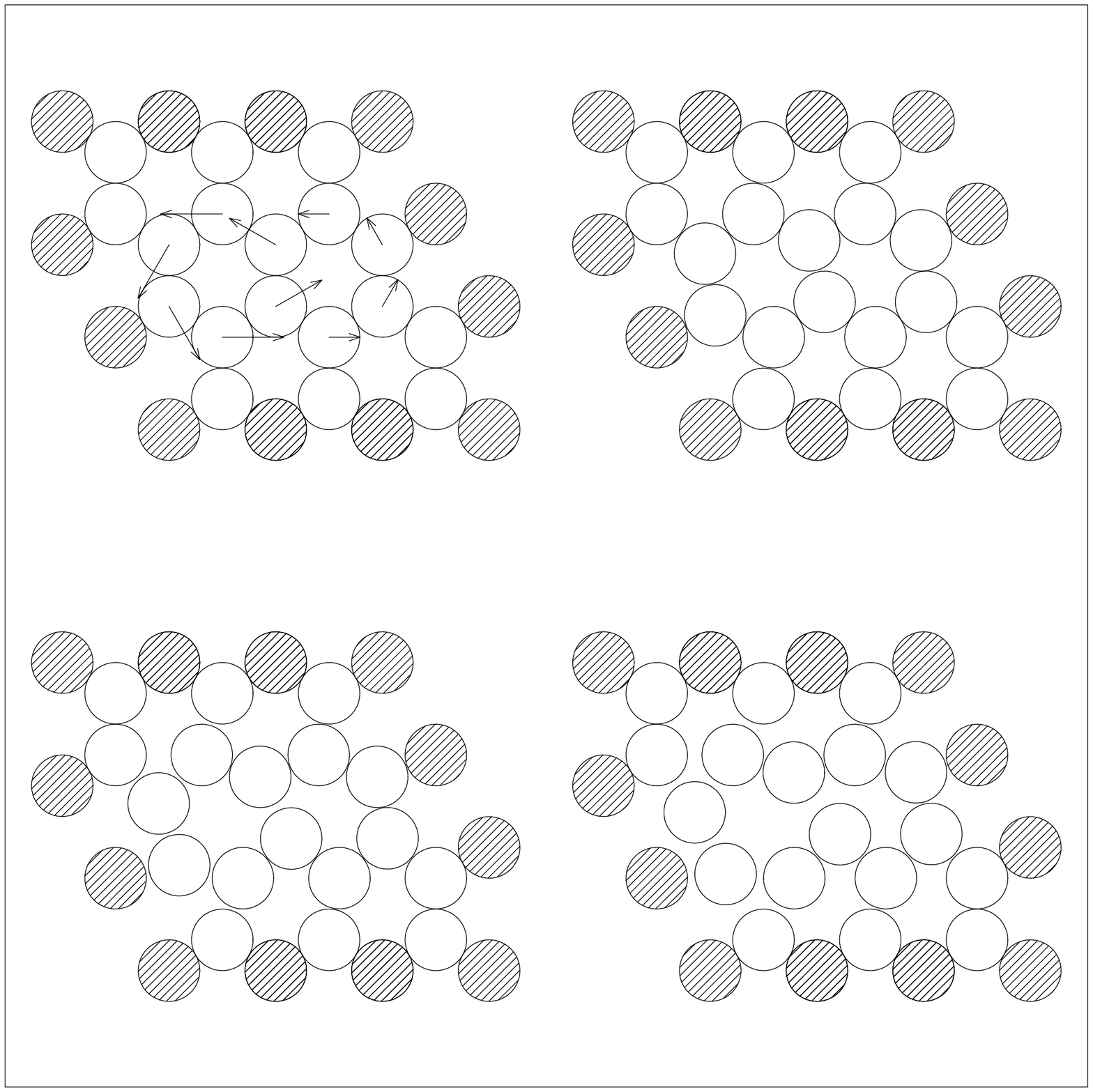}} \par}
   \caption{Donev \emph{et al.}}
\end{figure*}

\newpage
\begin{figure*}[!h]
   {\centering \resizebox*{0.75\columnwidth}{!}{\includegraphics{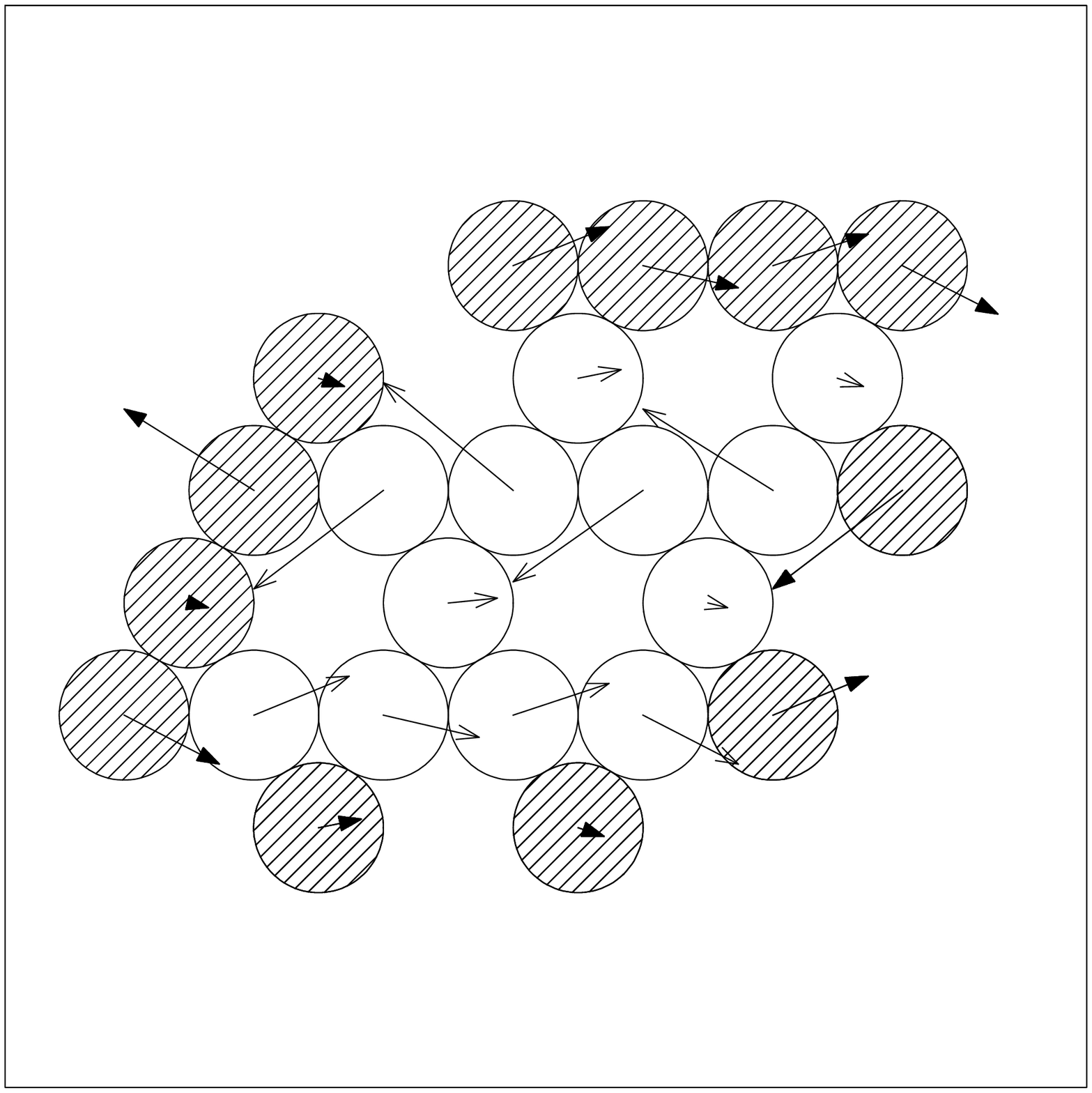}} \par}
   \caption{Donev \emph{et al.}}
\end{figure*}

\newpage
\begin{figure*}[!h]
   {\centering \resizebox*{1\textwidth}{!}{\includegraphics{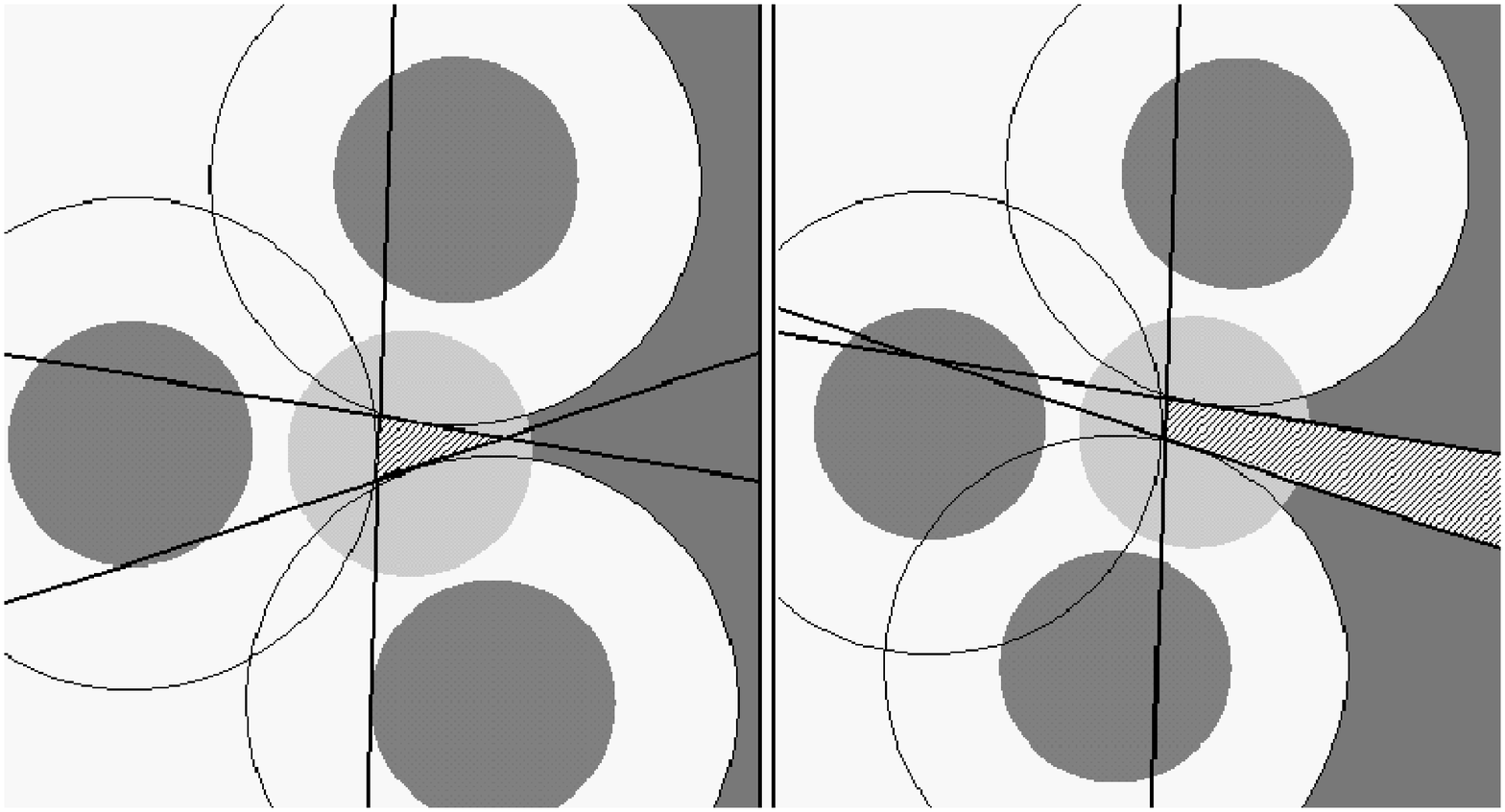}} \par}
   \caption{Donev \emph{et al.}}
\end{figure*}

\newpage
\begin{figure*}[!h]
   {\centering \resizebox*{6cm}{!}{\includegraphics{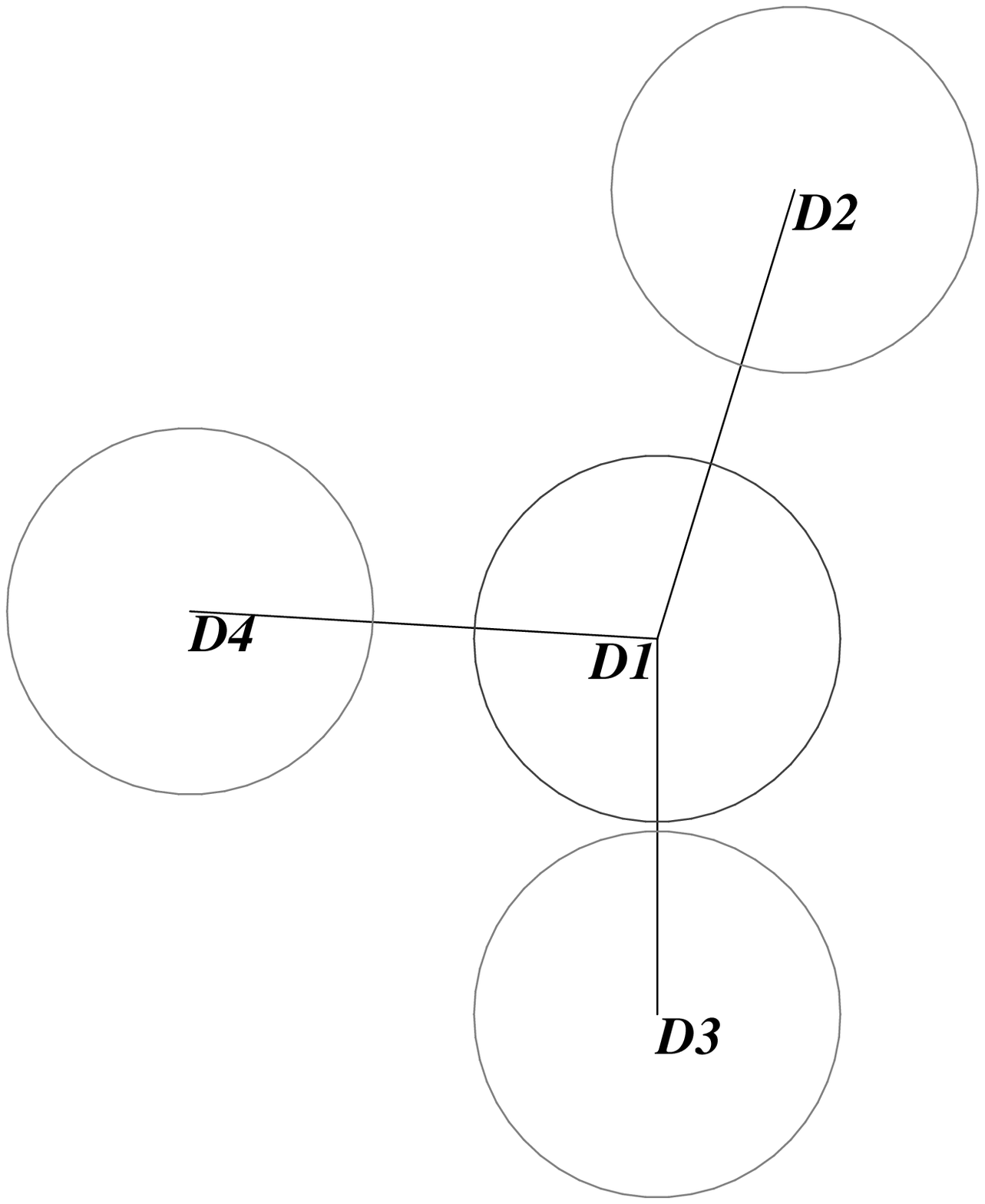}} \par}
   \caption{Donev \emph{et al.}}
\end{figure*}

\newpage
\begin{figure*}[!h]
   {\centering \resizebox*{1\textwidth}{!}{\includegraphics{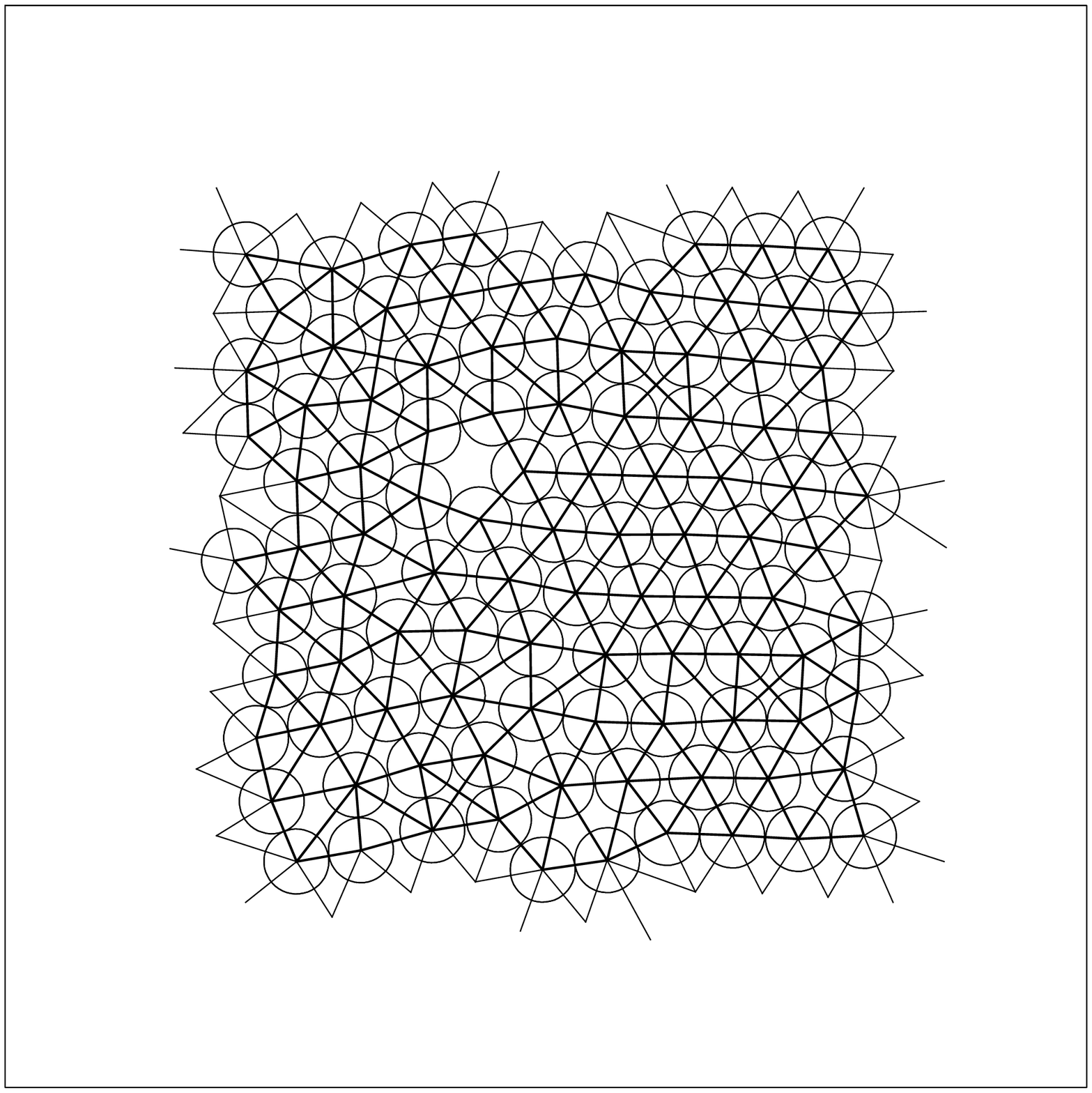}} \par}
   \caption{Donev \emph{et al.}}
\end{figure*}

\newpage
\begin{figure*}[!h]
   {\centering \resizebox*{!}{0.85\textheight}{\includegraphics{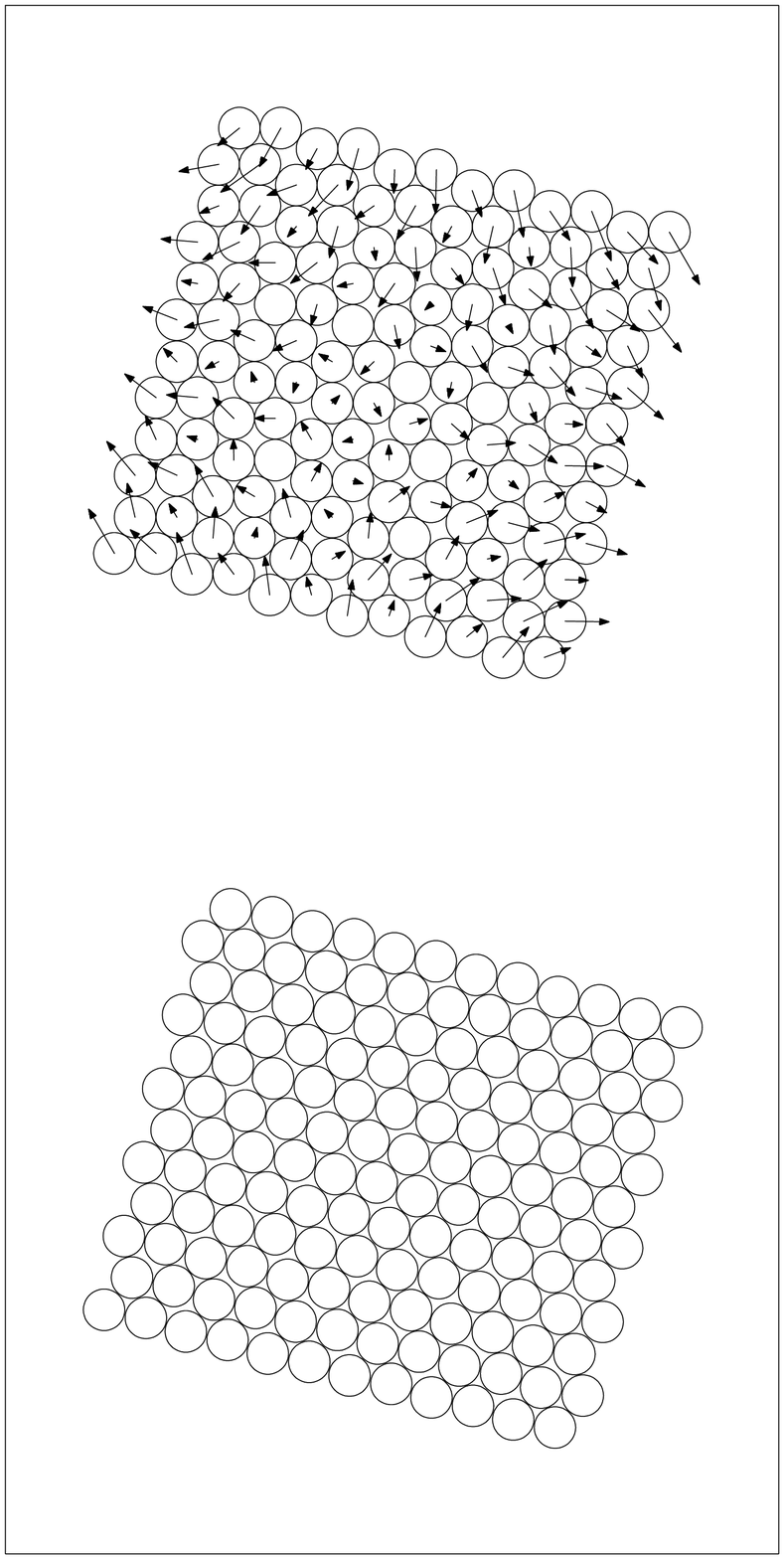}} \par}
   \caption{Donev \emph{et al.}}
\end{figure*}

\newpage
\begin{figure*}[!h]
   {\centering \resizebox*{0.4\textwidth}{!}{\includegraphics{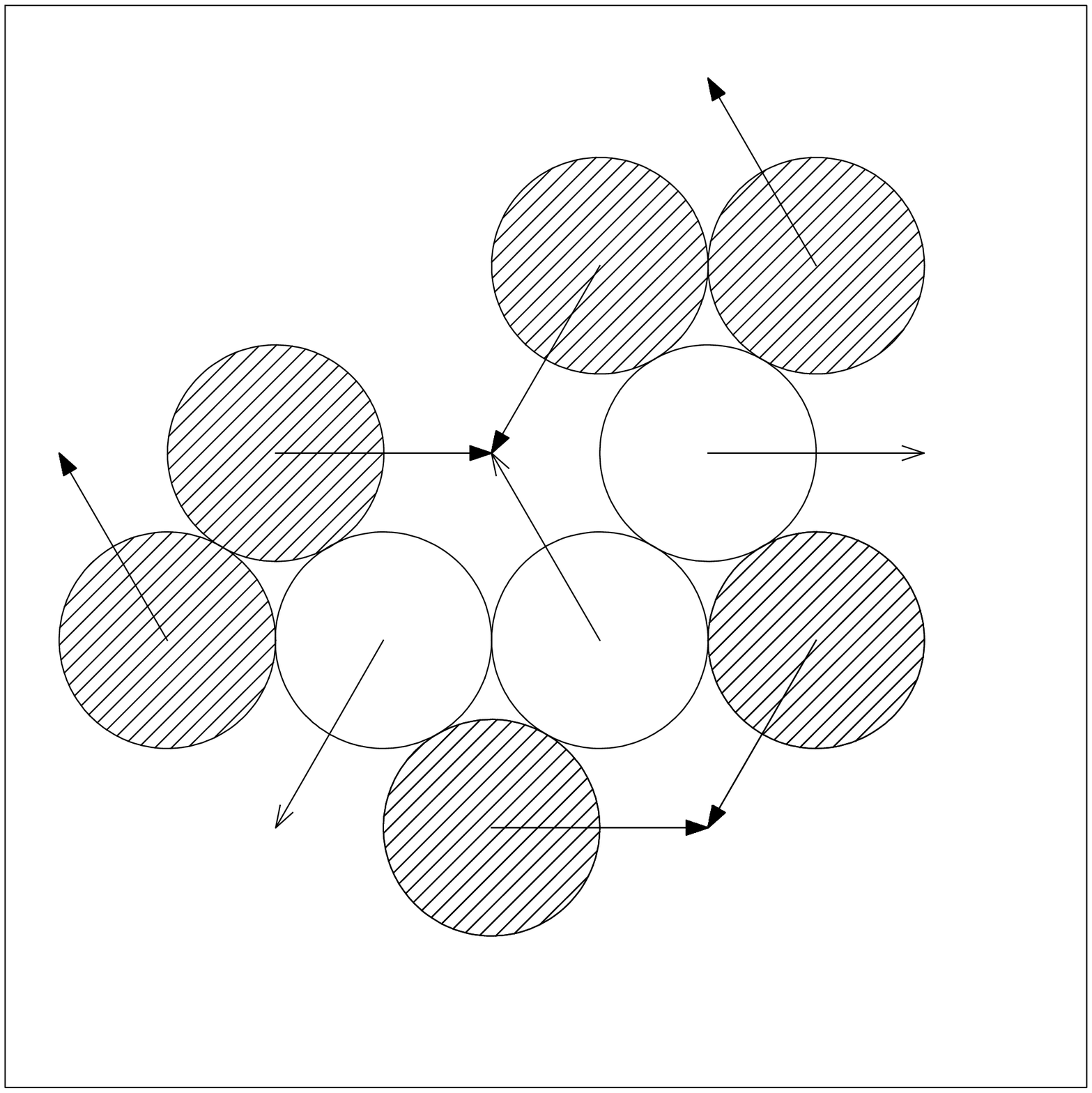}} \resizebox*{0.4\textwidth}{!}{\includegraphics{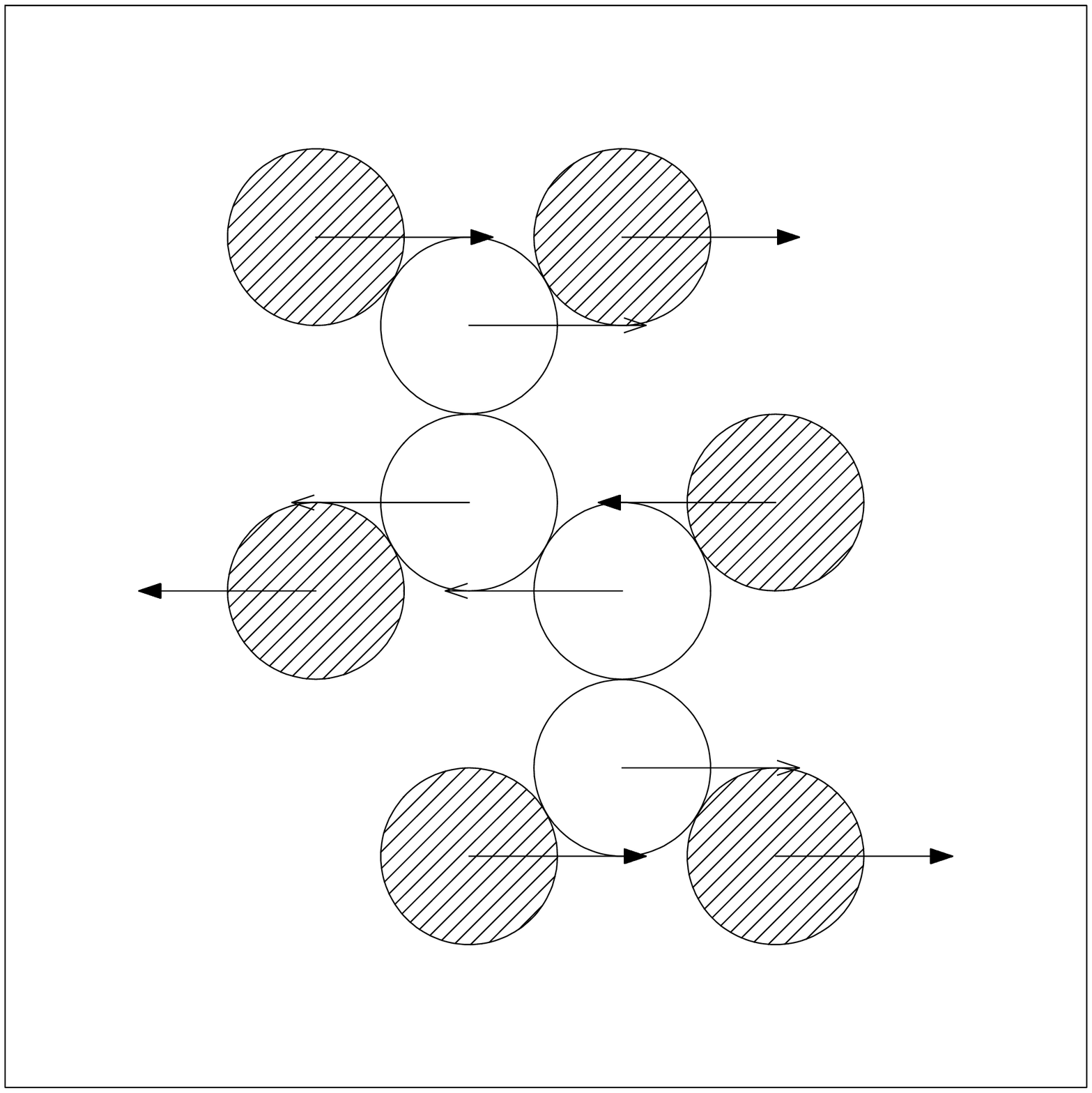}} \par}
   \caption{Donev \emph{et al.}}
\end{figure*}

\newpage
\begin{figure*}[!h]
   {\centering \resizebox*{1\textwidth}{!}{\includegraphics{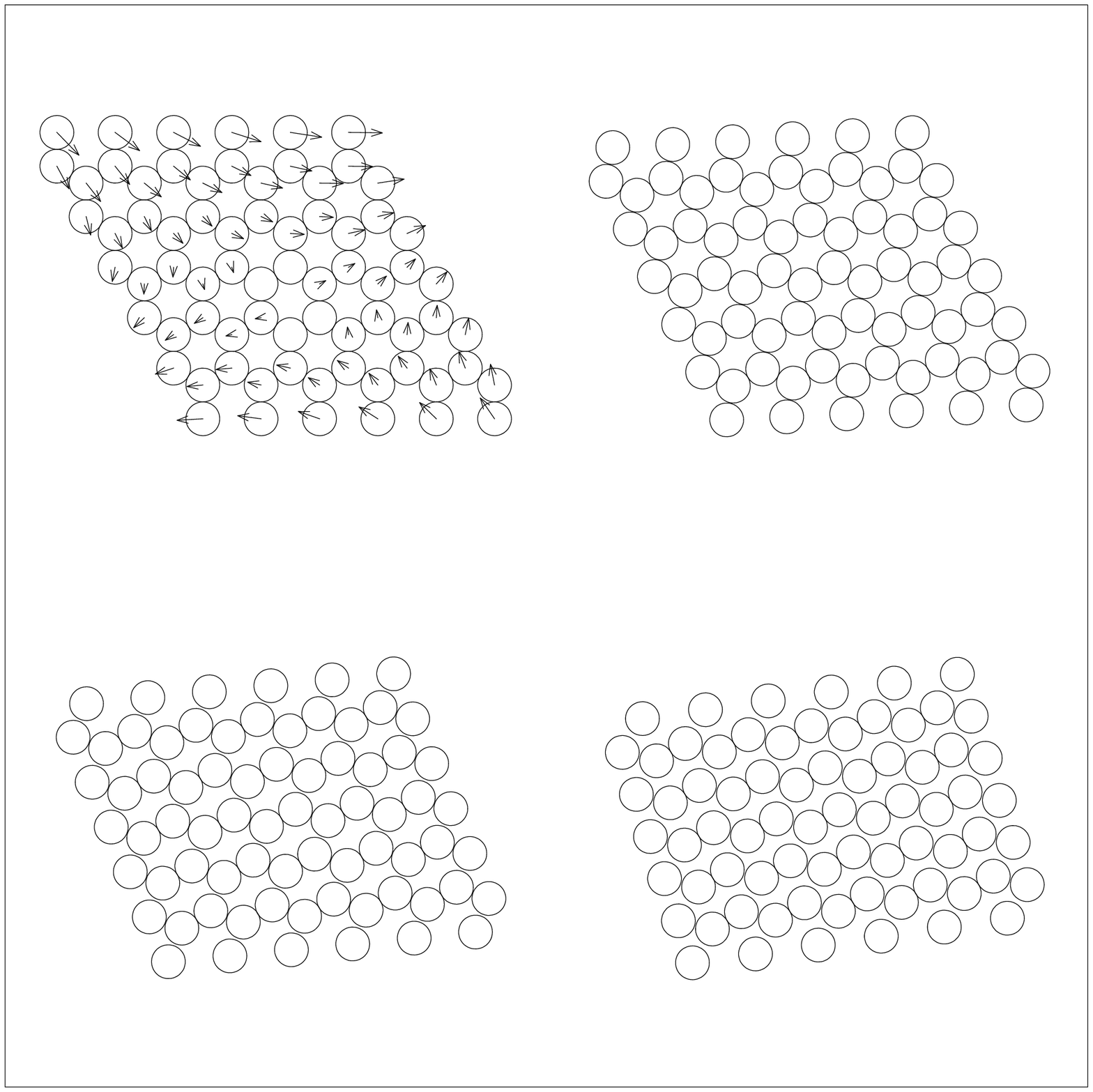}} \par}
   \caption{Donev \emph{et al.}}
\end{figure*}

\newpage
\begin{figure*}[!h]
   {\centering \resizebox*{0.49\textwidth}{!}{\includegraphics{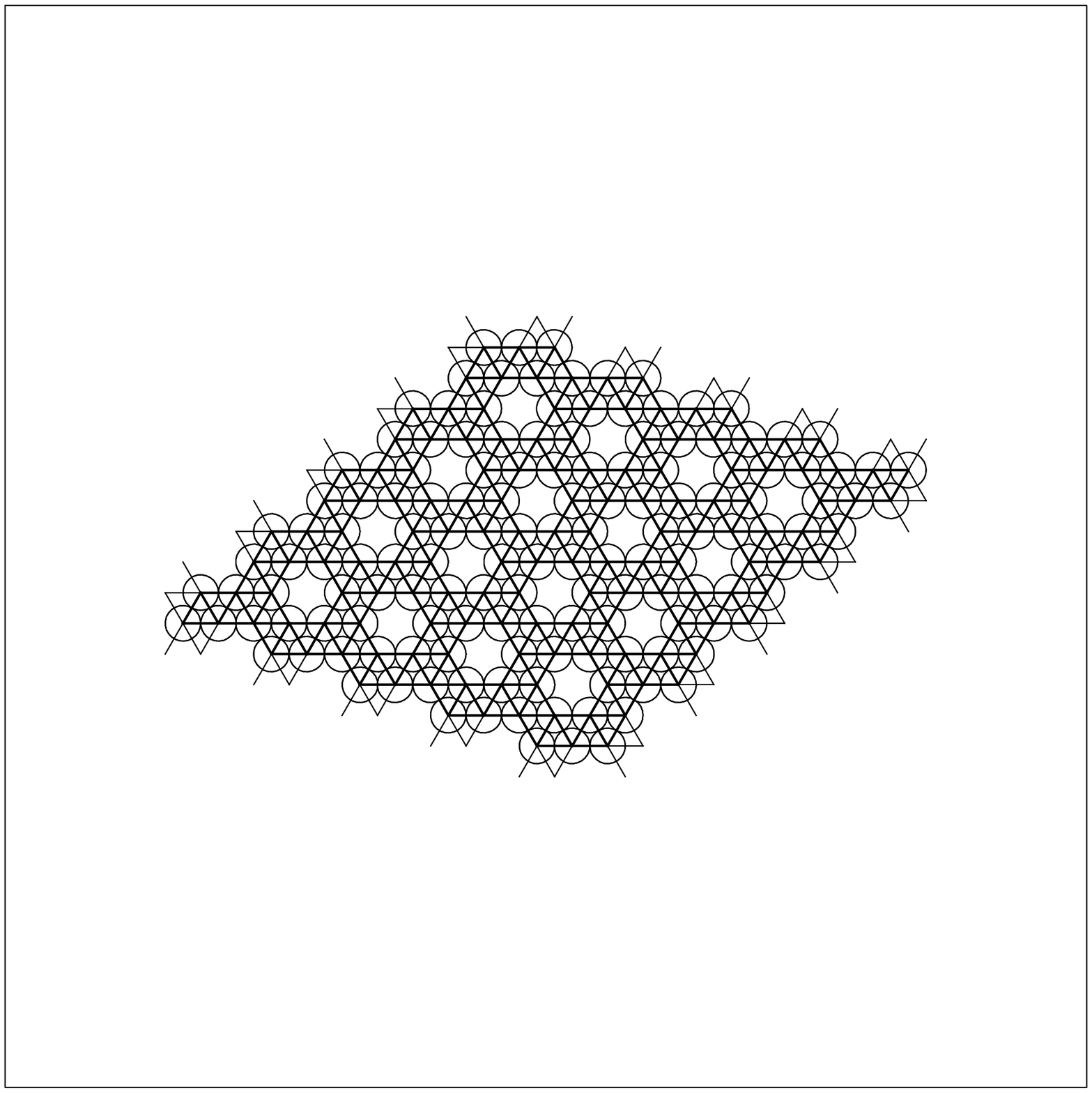}} \resizebox*{0.49\textwidth}{!}{\includegraphics{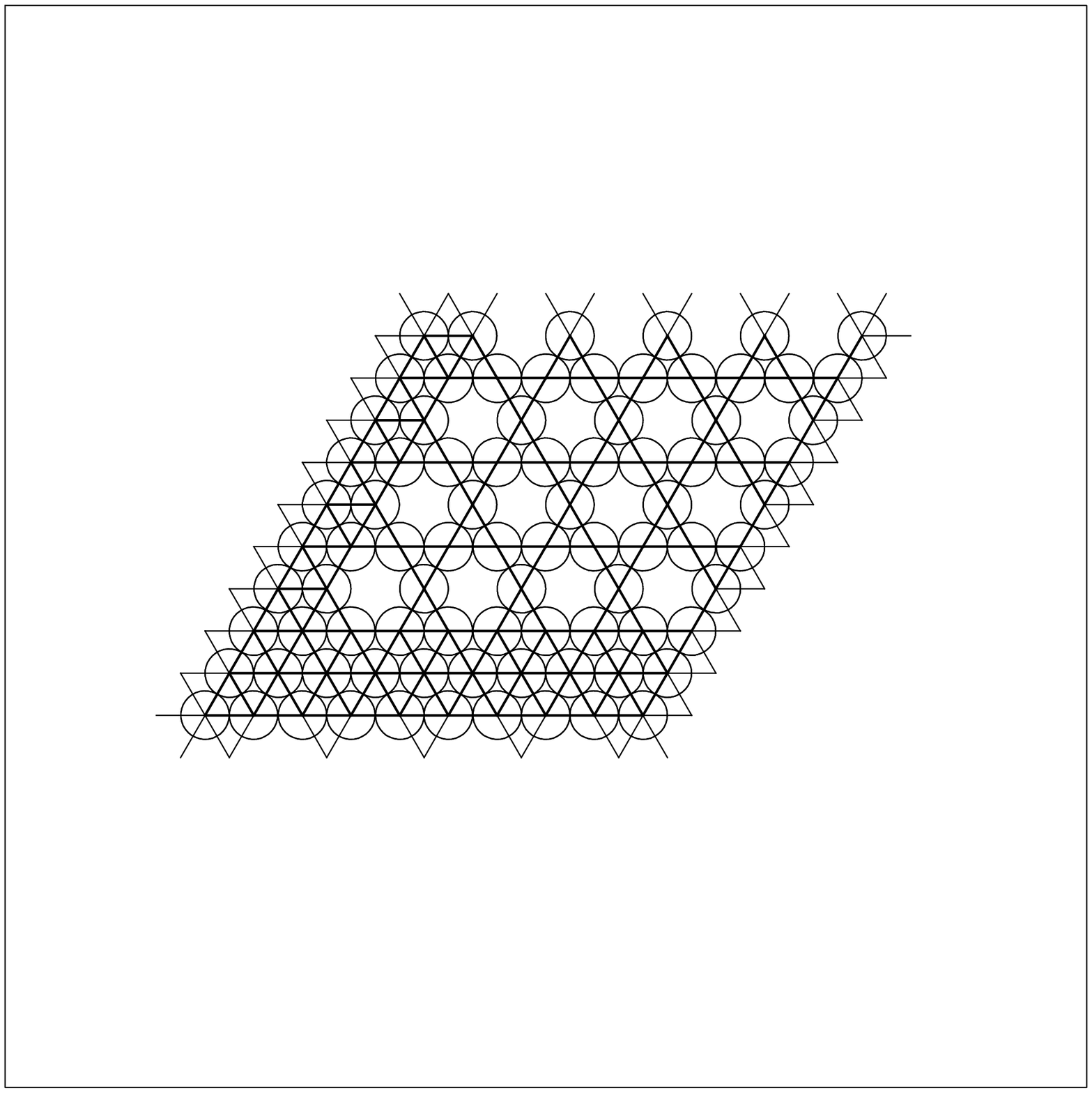}} \par}
   \caption{Donev \emph{et al.}}
\end{figure*}

\newpage
\begin{figure*}[!h]
   {\centering \resizebox*{1\textwidth}{!}{\includegraphics{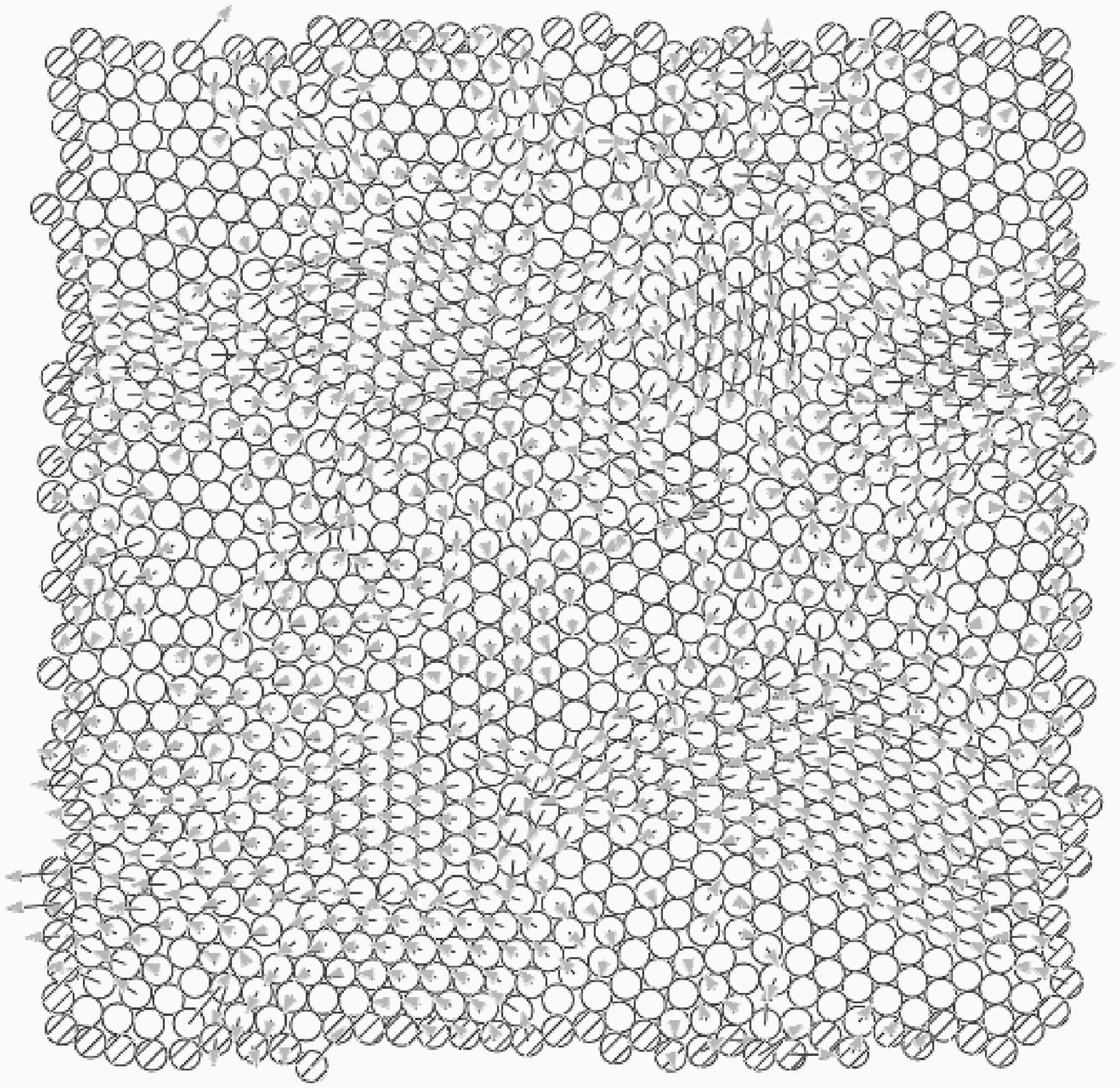}} \par}
   \caption{Donev \emph{et al.}}
\end{figure*}

\newpage
\begin{figure*}[!h]
   {\centering \resizebox*{1\textwidth}{!}{\includegraphics{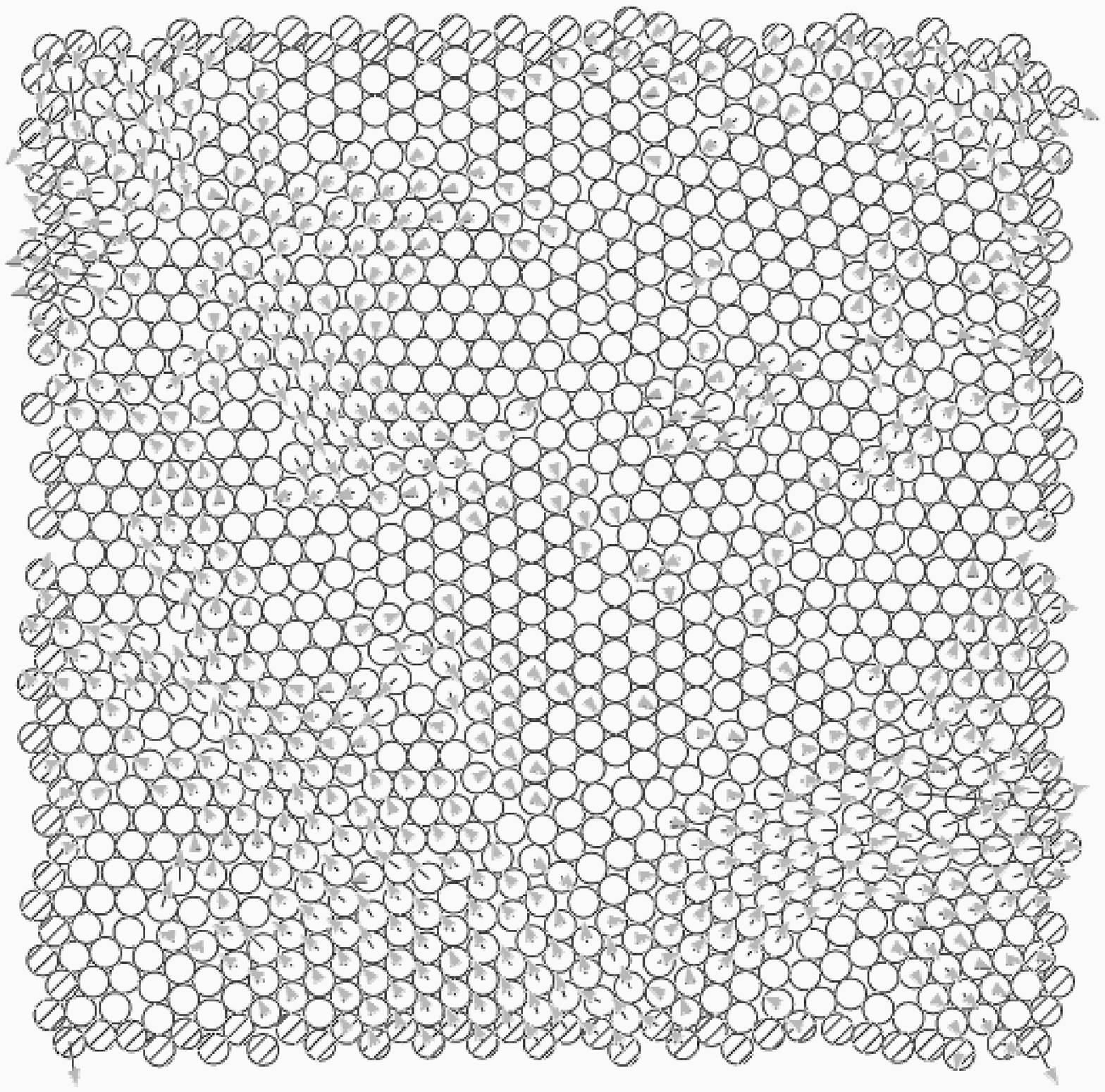}} \par}
   \caption{Donev \emph{et al.}}
\end{figure*}

\newpage
\begin{figure*}[!h]
   {\centering \resizebox*{1\textwidth}{!}{\includegraphics{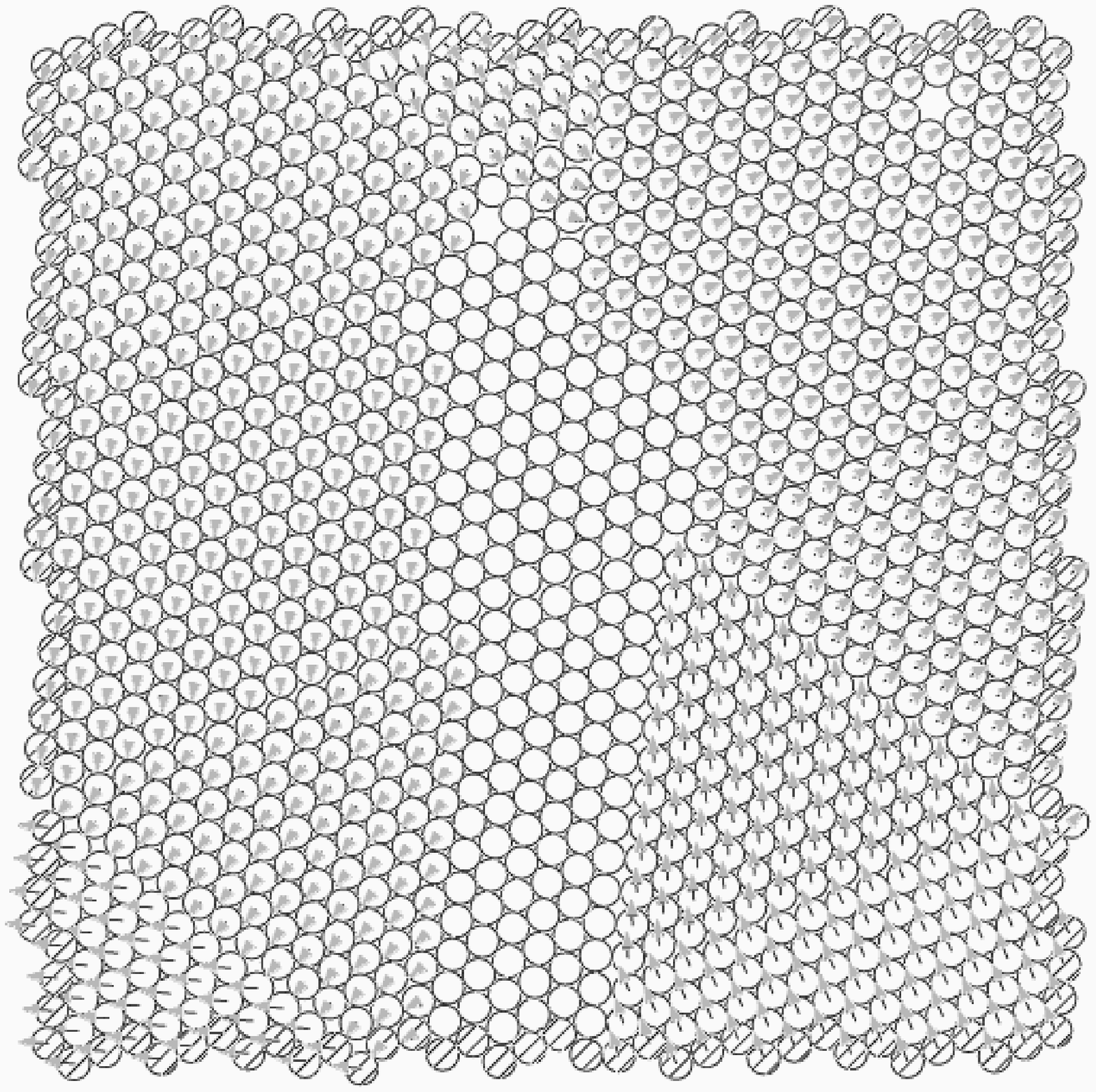}} \par}
   \caption{Donev \emph{et al.}}
\end{figure*}

\end{document}